\documentclass[%
 reprint,
 amsmath,amssymb,
 aps,
pra,
]{revtex4-1}
\usepackage[utf8]{inputenc}
\usepackage{amsmath}  
\usepackage{amsfonts} 
\usepackage{graphicx} 
\usepackage{physics}
\usepackage{url}
\usepackage{xcolor}
\usepackage[T1]{fontenc}
\usepackage[colorlinks, linkcolor=blue, citecolor=blue, urlcolor=blue, breaklinks=true]{hyperref}
\usepackage{float}

\preprint{APS/123-QED}

\begin{document}
\title{Higher-order Poincaré Spheres and Spatio-Spectral Poincaré Beams}

\author{R. Fickler}
\affiliation{Tampere University, Photonics Laboratory, Physics Unit, Tampere, FI-33720, Finland}
 \email{robert.fickler@tuni.fi}
\author{L. Kopf}
\affiliation{Tampere University, Photonics Laboratory, Physics Unit, Tampere, FI-33720, Finland}
\author{M. Ornigotti}%
\affiliation{Tampere University, Photonics Laboratory, Physics Unit, Tampere, FI-33720, Finland}

\date{\today}

\begin{abstract}
\noindent
The study of fundamental optics effects has been stimulated through the increasing ability to structure light in all its degrees of freedom (DOFs) in sophisticated but simple experimental settings. 
However, with such an increase in experimental capabilities, it has also become important to study theoretical descriptions for a more intuitive understanding of the underlying concepts.
Here, we introduce a visual representation of light that is structured in its transverse space, frequency, and polarization in the form of a higher-order Poincaré sphere and discuss interesting links to its fundamental counterpart.
We further leverage this connection to discuss and experimentally generate light possessing all possible polarization states across its spatio-spectral shape, which we term \emph{spatio-spectral Poincaré beams}.
By invoking all DOFs of light in the powerful description of higher-order Poincaré spheres, our work can pave the way for a deeper understanding and beneficial application of structured light as a powerful tool in optics.
\end{abstract}

\maketitle

Increasing the complexity of light fields by shaping their amplitude in space, time, and polarization, i.e., all their degrees of freedom (DOFs) is advancing research in various different fields. 
Driven by increasingly advanced technologies to control light, ideas from different optics fields have been adapted, combined, and extended leading to growing interest in fundamental studies of linear, nonlinear, and quantum optics, as well as applications in imaging, sensing, computing, and communications \cite{forbes2021structured, he2022towards}.
A lot of attention has been attributed to form complex spatial polarization states by combining polarization with different spatial mode structures, leading to so-called spatial vector beams \cite{chen2018vectorial}.
Spatial vector beams are used in applications such as tight focusing \cite{dorn2003sharper}, the study of complex entanglement pattern \cite{fickler2014quantum}, high-speed imaging schemes \cite{berg2015classically}, and the observation of three-dimensional polarization structures like knots and links \cite{larocque2018reconstructing}.
Mathematically, they have been described through a generalization of the fundamental Poincaré sphere (PS), the so-called higher-order PS for spatial vector beams \cite{Milione2011, holleczek2010poincare}, which enables an intuitive visual representation of these vectorial light fields and helps describing complex evolutions, as for example light fields obtained through a high-order Pancharatnam–Berry phase \cite{milione2012higher}.
Additionally, the concept of higher-order PS can be linked to different applications in communications \cite{ ndagano2017characterizing}, metamaterial design \cite{jiang2020single}, and extensions to more complex spatially structured light \cite{shen20202}.

A particularly interesting set of spatial vector beams describes light fields containing all possible polarization states, so-called Poincaré beams \cite{beckley2010full}. 
For such beams, the transverse polarization pattern can be mapped to the entire fundamental PS through a stereographic projection.
The patterns can further be linked to interesting topological features, such as complex polarization singularities \cite{dennis2002polarization},  
polarization Möbius stripes \cite{bauer2015observation}, and optical skyrmions in the classical \cite{shen2024optical} and quantum domain \cite{ornelas2024non} to highlight a few.

In this letter, we extend the idea of high-order PSs to states of light with a complex polarization structure over its spatial as well as spectral DOF, and present interesting links to its fundamental counterpart. We first demonstrate that single points on higher-order PSs of spatial as well as spectral vector beams can be linked to latitudes and meridians on the fundamental PS, respectively. We then introduce a more complex higher-order PS describing spatio-spectral vector beams, i.e., light with a spatially and spectrally imhomogenous polarization state. We show that such light fields can exhibit all possible polarization states over the spatio-spectral domain and how the patterns can be mapped onto the entire surface of the fundamental PS. As such, we term these states of light \emph{Spatio-Spectral Poincaré beams} (SSPBs). We then generate SSPBs using a simple experimental setup and verify that all polarization states of the PS are found using spatio-spectrally resolving polarization tomography.



The vectorial nature of light, i.e., its polarization, can be described as a two-dimensional state using the circular polarisation basis $\{\ket{R}$,$\ket{L}\}$, so that any state can be constructed as a superposition with 
\begin{eqnarray}
\ket{\Psi}=\cos\left( \frac{\theta}{2}\right)\ket{R} + \sin\left( \frac{\theta}{2}\right) e^{i\phi} \ket{L},
\label{eq:pol_state}
\end{eqnarray}
where $\theta\in[0,\pi]$ and $\phi\in[0,2\pi]$. 
These states are conveniently represented through the surface of the PS, where $\ket{R}$ and $\ket{L}$ are located on the north- and south-pole of the PS, respectively (see supplementary for more information). 
With a similar argument, one can also describe other DOFs of a paraxial light field, i.e., the transverse spatial \cite{padgett1999poincare, agarwal19992, dennis2017swings} and the spectral (temporal) \cite{brecht2015photon} DOF, or, in a more general framework, two-dimensional quantum states in the form of the Bloch sphere \cite{nielsen2010quantum}.

\begin{figure*}[ht] 
    \centering
    \includegraphics[width=0.9\linewidth]{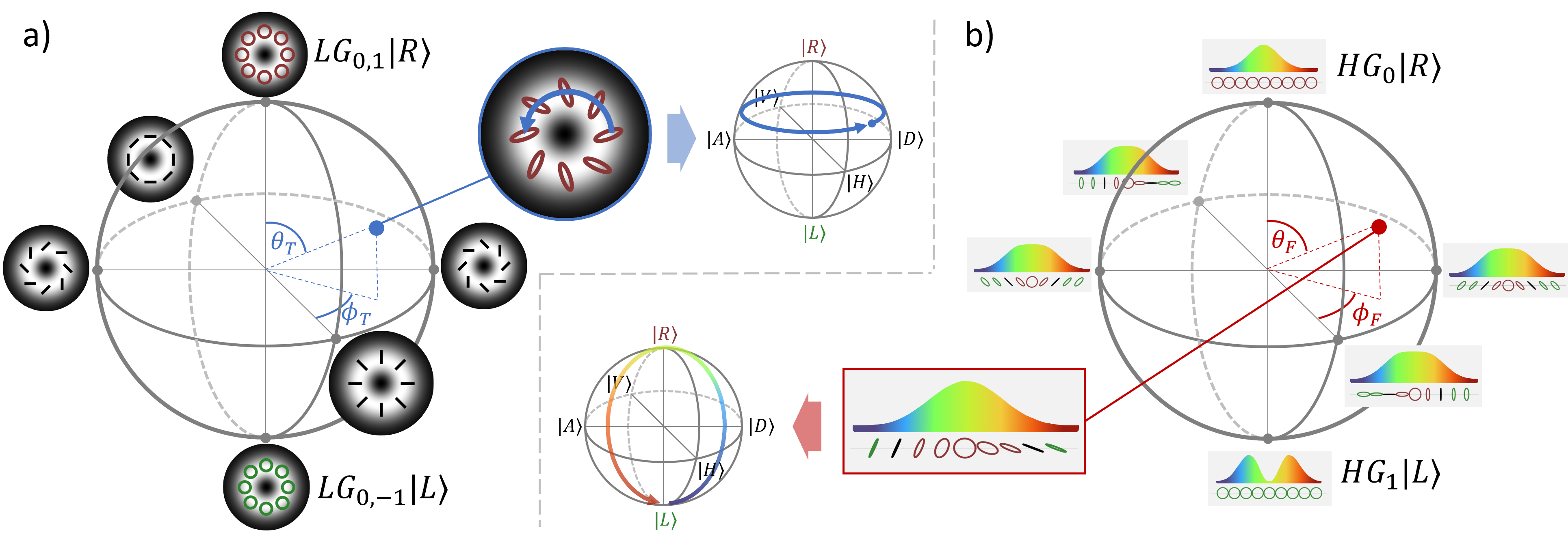}
    \caption{Higher-order Poincaré spheres (PS) in the space and frequency domain. 
    (a) Higher-order spatial PS constructed by two states at the poles with orthogonal circular polarization and two spatial Laguerre-Gaussian modes $LG_{0,\ell}$ with $\ell=\pm1$. 
    A single point on the sphere describes a beam with a transverse spatial polarization pattern, which itself can be visualized as a line along a latitude on the fundamental PS (see inset).
    The states on opposing sides are graphically depicted by their transverse intensity in grey scale overlaid by the polarization ellipses.
    (b) Higher-order spectral PS built by the two spectral Hermite-Gaussian modes $HG_n$ with $n=0,1$, each with the opposite circular polarization. 
    Here, a single point corresponds to a beam with a spectrally varying polarization pattern, whose distribution follows a meridian of the fundamental PS (see inset). 
    Graphically, the states are depicted by their spectral intensity in rainbow colors with the polarization ellipses underneath.    
    \(\ket{H}\), \(\ket{V}\), \(\ket{D}\), \(\ket{A}\), \(\ket{R}\), and \(\ket{L}\) stand for horizontal, vertical, diagonal, anti-diagonal, and left- and right-handed circular polarization states, respectively.
    Both inset examples, a) and b), are shown for $\theta=2\pi/3$ and $\phi=\pi/4$. For all polarization ellipses, linear polarization is depicted by black lines, right- and left-handed polarization by red and green ellipses, respectively.} 
    \label{fig:1}
\end{figure*}

When polarization is correlated with one of the other DOFs, a higher-order PS can be defined \cite{Milione2011,holleczek2010poincare}. 
In the spatial domain, any set of two orthogonal transverse spatial modes $T_{n}(\Vec{r})$ is sufficient, where  $n$ denotes the mode number, and \(\Vec{r}\) the transverse spatial coordinate.
Thus, Eq.~\eqref{eq:pol_state} is transformed to
\begin{equation}
\ket{\Psi}_T=T_{1}(\Vec{r})\cos\left( \frac{\theta_T}{2} \right)\ket{R} + T_{2}(\Vec{r}) \sin\left( \frac{\theta_T}{2} \right) e^{i\phi_T} \ket{L},    \label{eq:SpaVec_state}
\end{equation}
with $\theta_T\in[0,\pi]$ and $\phi_T\in[0,2\pi]$, as before.
The complex-valued spatial mode functions $T_{n}(\Vec{r})$ introduce a spatially varying polarisation pattern in the transverse plane, such that these beams are termed \emph{spatial vector beams}.
For the spatial higher-order PS, the two poles are described by two orthogonal polarization states, where each state has an orthogonal spatial mode function and where $\theta_T$ and $\phi_T$ span the whole surface of the sphere.
Note, that for $T_{1}=T_{2}$, the higher-order PS reduces to the fundamental PS.

Arguably, the most popular spatial vector beams are realized when the spatial mode functions are implemented by higher-order Laguerre-Gaussian (LG) modes, i.e., $T_{n}(\Vec{r})=LG_{p,\ell}=G(r)L_p^{\abs{\ell}}(\rho)e^{-i\ell\varphi}$, with $r$ being the radius, $\rho=2r^2/w_0^2$ with $w_0$ being the beam waist, $G(r)$ describing a Gaussian profile, $L_p^{\abs{\ell}}(\rho)$ are the generalized Laguerre polynomials, and  $\varphi$ the azimuthal angle (simplified description of the transverse field for clarity).
In the case of no radial structure ($L_0^{\abs{\ell}}(\rho)=1$), we can choose the two first order LG modes carrying opposite OAM ($\ell=\pm1$), such that $T_{\pm1}(\Vec{r})=G(r)e^{\pm i\varphi}$. 
Then, Eq.~\eqref{eq:SpaVec_state} becomes
\begin{equation}
\ket{\psi}_T=G(r)e^{i\varphi}\left[\cos\left(\frac{\theta_T}{2}\right)\ket{R} + \sin\left(\frac{\theta_T}{2}\right) e^{i(\phi_T-2\varphi)} \ket{L}\right].    \label{eq:SpaVec_state_LG}
\end{equation}
The resulting higher-order PS is shown in Fig.~\ref{fig:1}a, where the two poles of the sphere are described by two donut-shaped light fields with two orthogonal transverse uniform circular polarizations.
Any point on the higher-order PS, except the two poles, corresponds to one combination of $\{\theta_T,\phi_T\}$ forming a unique spatial vector beam.
It contains a transverse varying polarization state, where the angular coordinate $\varphi$ relates to different phases between the two circular polarization components and cycles through all possible phases twice.
Thus, the higher-order PS and the fundamental PS are linked through $\theta = \theta_T$ and $\phi = \phi_T+2\phi$. 
Interestingly, this results in each spatial vector beam containing all polarization states of a latitude (horizontal line) on the fundamental PS (see example and inset in Fig.~\ref{fig:1}a and supplementary material for a more detailed discussion).
\begin{figure*}[ht] 
    \centering
    \includegraphics[width=0.9\linewidth]{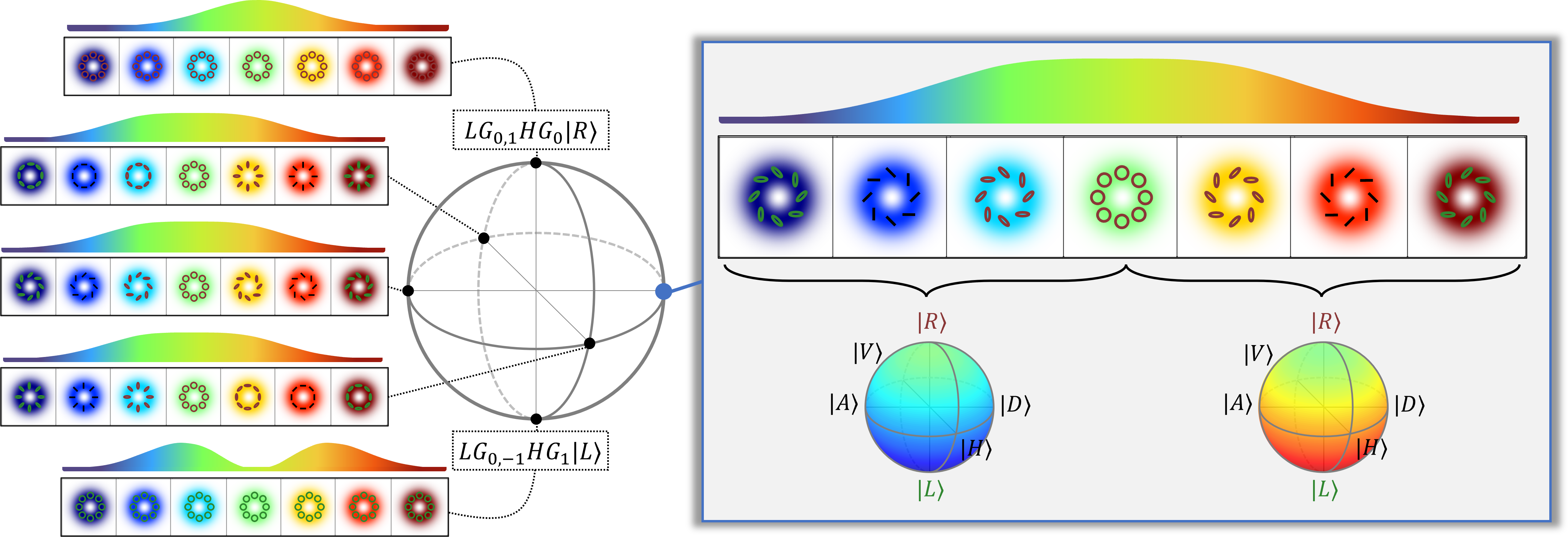}
    \caption{Spatio-spectral Poincaré sphere with the North (South) pole describing a right-circular (left-circular) polarization state with a Gaussian (first order Hermite-Gaussian) spectral mode and a transverse spatial Laguerre-Gaussian mode with $\ell=1$ ($\ell=-1$). 
    The inset on the right shows a single spatio-spectral Poincaré beam, where half of the spectrum covers the whole surface of a fundamental PS.
    Linear polarization is depicted by black lines, right- and left-handed circular polarization by red and green ellipses, respectively.} 
    \label{fig:2}
\end{figure*}

Although less studied, a similar vector beam can be constructed by combining polarization and the spectral or temporal DOF of light. 
As before, we can extend Eq.~\eqref{eq:pol_state} by including two orthogonal spectral mode functions $F_{n}(\omega)$ such that we obtain a so-called \emph{spectral vector beam} \cite{kopf2021spectral} of light described by the state
\begin{equation}
\ket{\Psi}_F=F_{1}(\omega)\cos\left(\frac{\theta_F}{2}\right)\ket{R} + F_{2}(\omega) \sin\left(\frac{\theta_F}{2}\right) e^{i\phi_F} \ket{L}, 
\label{eq:SpeVec_state}
\end{equation}
with $\omega$ being the angular frequency of the light field, $\theta_F\in[0,\pi]$, and $\phi_F\in[0,2\pi]$.
The states are forming the surface of a higher-order spectral PS, where the poles are describing two orthogonal circular polarization states, each having a different orthogonal spectral mode.
As before, for $F_1=F_2$, the higher-order PS reduces to the fundamental one. 
Note, that although here the higher-order PS is discussed in the spectral domain, all results hold also for the temporal domain as both are linked via the Fourier relation. 

Commonly, the spectral mode functions of choice are Hermite-Gaussian (HG) spectral modes with $F_{n}(\omega)=HG_n=G(\omega)H_n(\Omega)$, where $G(\omega)$ describes a Gaussian profile, $H_n$ are the Hermite polynomials, and $\Omega=(\omega_0-\omega)/\sigma$ with $\sigma$ labelling the width of the spectrum and $\omega_0$ its central frequency \cite{brecht2015photon}.  
For the two lowest order modes, i.e., the first two Hermite polynomials $H_0(x)=1$ and $H_1(x)=2x$, we obtain $F_1(\omega)=G(\omega)$ and $F_2=G(\omega)2\Omega$ and Eq.~\eqref{eq:SpeVec_state} can be rewritten as 
\begin{align}
\ket{\Psi}_F=& G(\omega)\Omega' \left[\frac{1}{\Omega'}\cos\left(\frac{\theta_F}{2}\right)\ket{R}+ \Omega' \sin\left(\frac{\theta_F}{2}\right) e^{i\phi_F} \ket{L}\right],    \label{eq:SpeVec_state_HG}
\end{align}
with $\Omega'=\sqrt{2\Omega}$.
The resulting higher-order PS is shown in Fig.~\ref{fig:1}b, where all but the two states on the poles have a polarization that varies depending on the frequency component.
In contrast to the previously introduced spatial vector beams, these spectral vector beams show a different connection to the fundamental PS.
Eq.~\eqref{eq:SpeVec_state_HG} shows that for $\Omega'=0$, i.e., $\omega=\omega_0$ the beam is in state \(\ket{L}\).
For $\Omega'\rightarrow\pm\infty$ with a fixed phase $\pm\phi_F$ between the right- and left-handed component, the polarization continuously evolves to \(\ket{R}\).
The correct mapping of these beams onto the fundamental PS is then given by $\phi=\phi_F$ and $\theta=\arctan\left(\Omega'\,\tan\theta_F/2\right)$.
Thus, any point $\{\theta_F,\phi_F\}$ on the higher-order PS contains all polarization states of a meridian (vertical line) on the fundamental PS (see inset in Fig.~\ref{fig:1}b, and
 the supplementary material).

To fully describe a complex vectorial light field in space and spectrum (time), we combine all DOFs, such that one can write
\begin{align}
\ket{\Psi}_{TF}=&T_1(\Vec{r})F_{1}(\omega)\cos\left(\frac{\theta_{TF}}{2}\right)\ket{R} \nonumber\\
+&T_2(\Vec{r})F_{2}(\omega)\sin\left(\frac{\theta_{TF}}{2} \right) e^{i\phi_F}\ket{L}.    
\label{eq:SpaSpeVec_state}
\end{align}
Because the polarization state changes across the transverse beam profile as well as its frequency spectrum, these beams can be termed \emph{spatio-spectral vector beams} \cite{kopf2023correlating}.
As before, for $T_1=T_2$ and $F_1=F_2$ the description reduces to simple homogeneous polarization states. If only one of the two cases hold, the reduction instead leads to a spatial or spectral vector beam.

In the following, we use $T_{1,2}(\Vec{r})=LG_{0,\pm1}$ and $F_{1,2}(\omega)=HG_{0,1}$, to align with the examples of higher-order PSs described above.
As before, it is possible to define a more complex higher-order PS (see Fig. \ref{fig:2}), where the two poles have a uniform polarization.
However, all other states now contain all possible polarization states of the fundamental PS when sampled across space and spectrum, i.e., they are spatio-spectral Poincaré beams (SSPB).
Analogue to spatial and spectral vector beams, all SSPBs states, excluding those on the poles, have a frequency dependent amplitude ratio between their \(\ket{L}\) and \(\ket{R}\) contributions.
At the same time, the relative phase between \(\ket{L}\) and \(\ket{R}\) cycles from 0 to 2$\pi$ twice dependent on the transverse angle $\varphi$.
In other words, the polarization pattern of such a beam is a complex stereographic projection of the PS onto the three-dimensional space of the spatio-spectral domain (see supplementary information for more details).
\begin{figure}[ht] 
    \centering
    \includegraphics[width=0.9\linewidth]{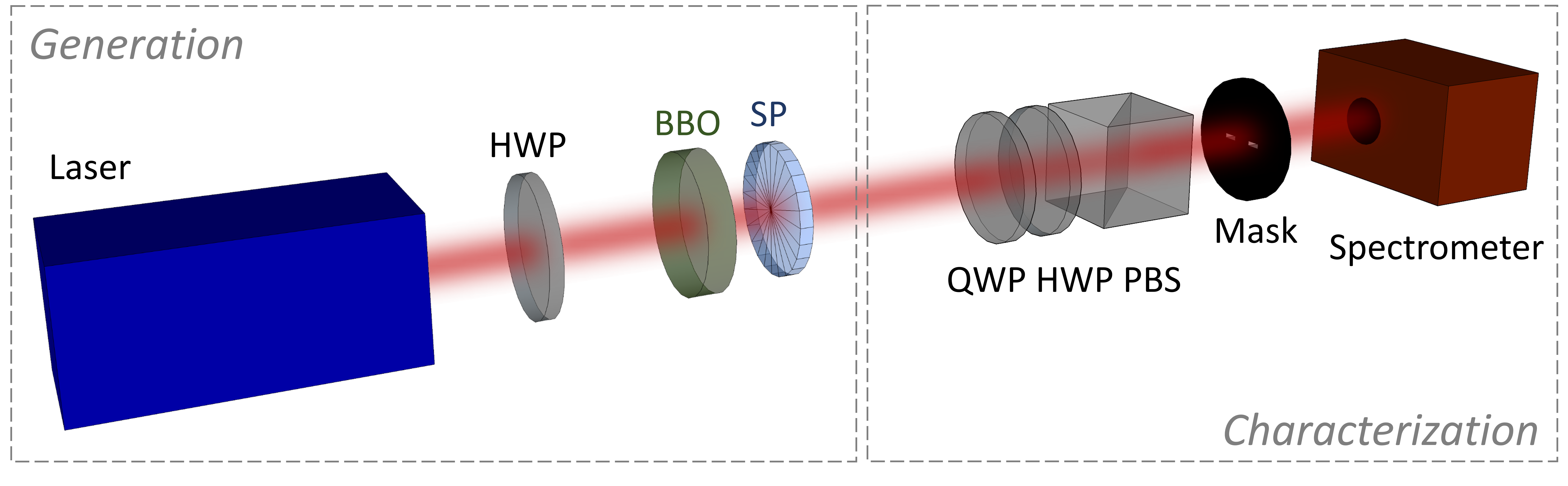}
    \caption{Scheme to generate a Spatio-Spectral Poincaré beam (SSPB). By using a half-wave plate (HWP), a beam from a fs-laser is diagonally polarized with respect to the optical axis of the birefringence crystal (BBO). The generated spectral vector beam is then sent through an S-vortex plate (SP) to generate a SSPB.    
    The SSPB is characterized through spatio-spectrally resolved polarization tomography using polarization optics (quarter-wave plate (QWP), HWP, polarizing beam splitter (PBS)), a spatial slit mask, and a spectrometer.} 
    \label{fig:3}
\end{figure}
Let's visualize this feature by analysing the exemplary state on the equator of the higher-order PS with $\{\theta_{TF}=\pi/2,\phi_{TF}=\pi/2\}$ (inset in Fig.~\ref{fig:2}).
The corresponding points $P$ on the fundamental PS are given by
\begin{equation}
P_{\frac{\pi}{2},\frac{\pi}{2}}=(\sin\arctan\Omega')\{\sin\,2\varphi,\cos\,2\varphi,1\}.
\end{equation}
If we fix the value of $\Omega'$, 
$P_{\frac{\pi}{2},\frac{\pi}{2}}$ evolves as the angle $\varphi$ changes from $0$ to $2\pi$ and creates a doubly covered great circle in the equatorial plane, 
If we instead fix a value of $\varphi$ and vary the frequency $\Omega'$ from $-\infty$ to $+\infty$, $P_{\frac{\pi}{2},\frac{\pi}{2}}$ describes a great circle in the meridian plane, i.e., from the South pole to the North pole. 
Notice, that as $\omega\rightarrow\pm\infty$, $\arctan\Omega\rightarrow\pm\pi/2$, which is equivalent to $\sin\arctan\Omega\rightarrow\,0,\pi.$

\begin{figure}[b] 
    \centering
    \includegraphics[width=0.9\linewidth]{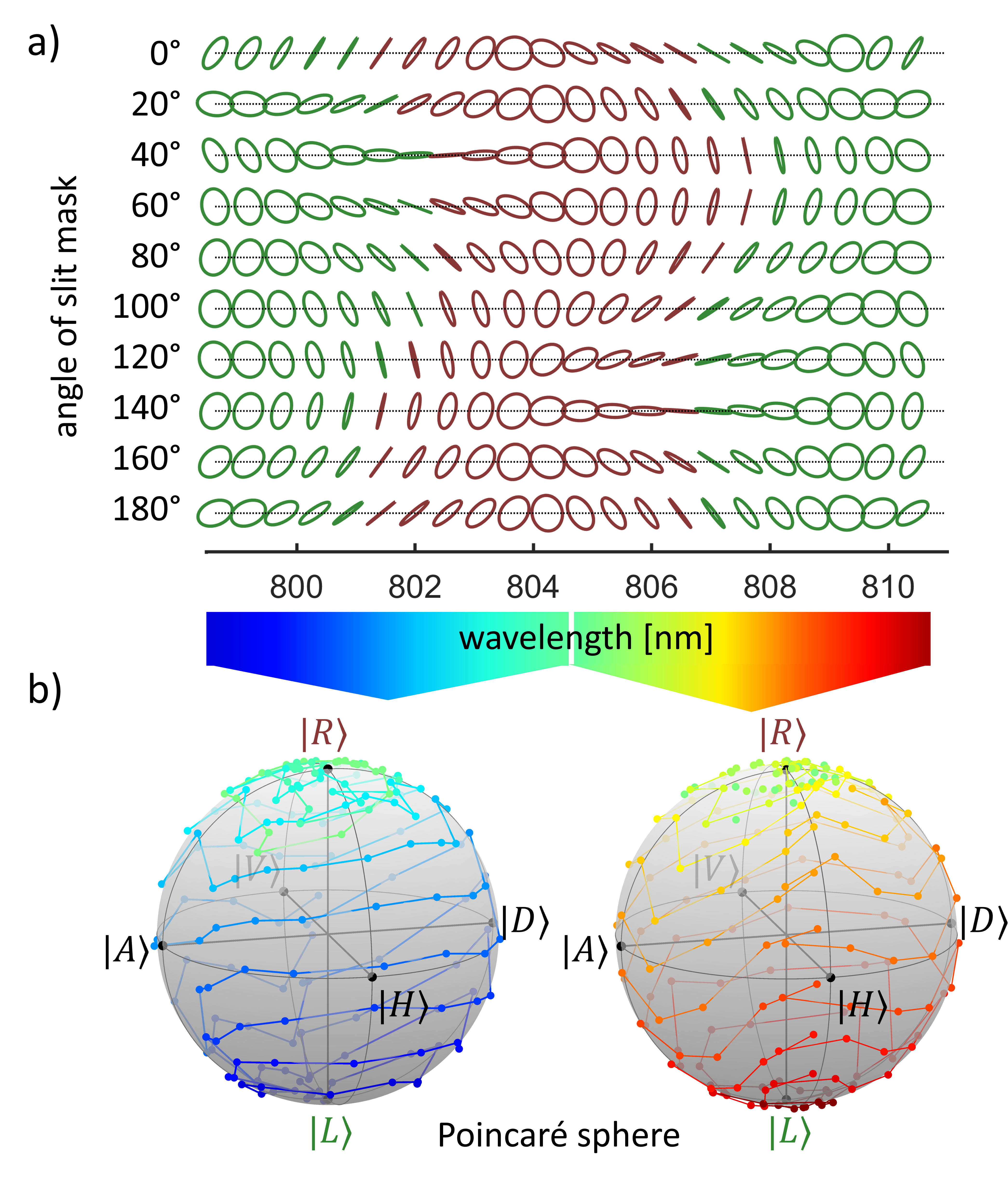}
    \caption{Experimentally reconstructed polarization pattern of a Spatio-Spectral Poincaré beam. 
    a) Reconstructed polarization ellipses depending on the angle of the slit mask and wavelength. Right- and left-handed polarizations are depicted by red and green ellipses, respectively. 
    b) All measured spatio-spectral polarization states on their respective position on the fundamental PS.
    All data points measured at the same wavelength for different spatial mask angles are connected by a line and colored according to their wavelength. Ideally, the lines are latitudes, however, measurement errors lead to slight deformations. } 
    \label{fig:4}
\end{figure}

There are well-known experimental techniques generate SSPB's by using the first higher-order LG modes in space \cite{bolduc2013exact} and HG modes in frequency \cite{weiner2011ultrafast}
However, there is a surprisingly simple method to approximate them well by using only a two optical elements.
In the spectral domain, it was recently shown that a birefringence material can be used to create a polarization-dependent temporal splitting of a laser pulse and thereby producing a spectral varying polarization pattern, i.e., a spectral vector beam \cite{kopf2021spectral}. 
This technique does not generate two orthogonal spectral modes, i.e. the overlap between the two resulting spectral distributions is small but non-vanishing, however, over a certain wavelength range, the obtained state approximately translates to Eq. \eqref{eq:SpeVec_state_HG} with $\{\theta_{F}=\pi/2,\phi_{F}=\pi/2\}$.
In the spatial domain, the key component to generate a polarization-dependent azimuthal phase distribution between the two LG modes of opposite handedness is a segmented half-wave plate know as an S-vortex plate \cite{biener2002formation}.
More details on these approximations along with the individual experimental generation of a spectral and spatial vector beam can be found in the supplementary.

In our experiment, we use Fourier-limited laser pulses with approximately 220\,fs pulse duration centered at a wavelength of around 804.5\,nm. 
At first, we generate a spectral vector beam by modulating the beams polarization with a half-wave plate which is aligned at 45$^\circ$ with respect to the optical axis of a 2\,mm thick birefringence BaB$_2$O$_4$ crystal (BBO) with a cut-angle of 23.4$^\circ$. 
After propagation through the BBO crystal, the laser pulses are coherently split into two trailing pulses of half its initial amplitude, which results in the required spectral polarization pattern.
Subsequently, we sent the laser through an S-vortex plate, which imprints the polarization-dependent OAM ($\ell=\pm1$) thereby completing the generation scheme.
See Fig.~\ref{fig:3} for a simplified scheme of the setup.

To characterize the generated SSPB, we perform spatio-spectrally-resolved polarization tomography.
We first filter the beam for different polarizations with a set of wave plates and a polarizing beam splitter and for angular positions with a slit mask (with an opening angle of approximately 14$^\circ$) and then measure its wavelength-resolved intensity with a spectrometer. 
From these measurements, we reconstruct the polarization ellipses for different wavelengths and angular positions of the slit mask as shown in Fig.~\ref{fig:4}a (see supplementary information for more details).
We find, that the reconstructed polarizations states nicely match the expected patterns described above, despite some experimental imperfections, such as the non-vanishing overlap of the spectral modes, and small optical misalignments.  
To show that the spatio-spectral polarization patterns of the generated beams indeed cover the full PS, we show the measured polarization states on the surface of the fundamental PS and, again, find a nice agreement apart from minor experimental imperfections (Fig.~\ref{fig:4}b).

In conclusion, we introduce an extension to the concept of higher-order PSs by including all DOFs of light and 
discuss how light can exhibit every possible polarization state. 
Furthermore, we experimentally generate a SSPB and 
verify that its polarization ellipses indeed cover the whole PS. 
Following this initial study, a multitude of different research opportunities could arise.
A direct next step can be to investigate the SSPBs in connection to the Pancharatnam–Berry phase, for which the introduced higher-order PS will be essential \cite{milione2012higher}. 
Another promising path will be to examine SSPBs in terms of complex polarization topologies, which have mostly been studied in the spatial  domain \cite{larocque2018reconstructing, shen2024optical, dennis2002polarization}, and see how they relate to polarization states of polychromatic light \cite{pisanty2019knotting, sugic2020knotted}. Moreover, having access to a higher-order PS representation for all DOFs of light fields could provide new insight into the geometrical properties of multipartite systems, as higher-order PSs allow an easier representation of such systems. This scheme, in fact, can be easily generalised to a system possessing $N$ DOFs, thus giving access to a geometrical representation of multipartite systems beyond the present schemes based on fibrations of spheres, which are limited to tripartite states \cite{fibration1,fibration2}. 
Finally, as all DOFs of the light field are strongly correlated in SSPBs, they can also serve as a valuable system to explore the similarities of classical and quantum non-separable states of light \cite{kopf2023correlating}.
Here, an extension of the recently discussed connection between spatial vector beams and entangled photon pairs \cite{peters2023spatially, ornelas2024non} could be extended to an analogy between SSPBs and tripartite quantum entangled system leading to interesting non-local features. 

\begin{acknowledgments}
R. F.  acknowledges the support of the Research Council of Finland through the Academy Research Fellowship (decision 332399). 
L. K. acknowledges the support by the Vilho, Yrjö and Kalle Väisälä Foundation of the Finnish Academy of Science and Letters.
All authors acknowledge the support of the Research Council of Finland through the Photonics Research and Innovation Flagship (PREIN - decision 320165).
\end{acknowledgments}

\bibliography{Biblio.bib}

\begin{thebibliography}{36}%
\makeatletter
\providecommand \@ifxundefined [1]{%
 \@ifx{#1\undefined}
}%
\providecommand \@ifnum [1]{%
 \ifnum #1\expandafter \@firstoftwo
 \else \expandafter \@secondoftwo
 \fi
}%
\providecommand \@ifx [1]{%
 \ifx #1\expandafter \@firstoftwo
 \else \expandafter \@secondoftwo
 \fi
}%
\providecommand \natexlab [1]{#1}%
\providecommand \enquote  [1]{``#1''}%
\providecommand \bibnamefont  [1]{#1}%
\providecommand \bibfnamefont [1]{#1}%
\providecommand \citenamefont [1]{#1}%
\providecommand \href@noop [0]{\@secondoftwo}%
\providecommand \href [0]{\begingroup \@sanitize@url \@href}%
\providecommand \@href[1]{\@@startlink{#1}\@@href}%
\providecommand \@@href[1]{\endgroup#1\@@endlink}%
\providecommand \@sanitize@url [0]{\catcode `\\12\catcode `\$12\catcode
  `\&12\catcode `\#12\catcode `\^12\catcode `\_12\catcode `\%12\relax}%
\providecommand \@@startlink[1]{}%
\providecommand \@@endlink[0]{}%
\providecommand \url  [0]{\begingroup\@sanitize@url \@url }%
\providecommand \@url [1]{\endgroup\@href {#1}{\urlprefix }}%
\providecommand \urlprefix  [0]{URL }%
\providecommand \Eprint [0]{\href }%
\providecommand \doibase [0]{http://dx.doi.org/}%
\providecommand \selectlanguage [0]{\@gobble}%
\providecommand \bibinfo  [0]{\@secondoftwo}%
\providecommand \bibfield  [0]{\@secondoftwo}%
\providecommand \translation [1]{[#1]}%
\providecommand \BibitemOpen [0]{}%
\providecommand \bibitemStop [0]{}%
\providecommand \bibitemNoStop [0]{.\EOS\space}%
\providecommand \EOS [0]{\spacefactor3000\relax}%
\providecommand \BibitemShut  [1]{\csname bibitem#1\endcsname}%
\let\auto@bib@innerbib\@empty
\bibitem [{\citenamefont {Forbes}\ \emph {et~al.}(2021)\citenamefont {Forbes},
  \citenamefont {de~Oliveira},\ and\ \citenamefont
  {Dennis}}]{forbes2021structured}%
  \BibitemOpen
  \bibfield  {author} {\bibinfo {author} {\bibfnamefont {A.}~\bibnamefont
  {Forbes}}, \bibinfo {author} {\bibfnamefont {M.}~\bibnamefont {de~Oliveira}},
  \ and\ \bibinfo {author} {\bibfnamefont {M.~R.}\ \bibnamefont {Dennis}},\
  }\href@noop {} {\bibfield  {journal} {\bibinfo  {journal} {Nature Photonics}\
  }\textbf {\bibinfo {volume} {15}},\ \bibinfo {pages} {253} (\bibinfo {year}
  {2021})}\BibitemShut {NoStop}%
\bibitem [{\citenamefont {He}\ \emph {et~al.}(2022)\citenamefont {He},
  \citenamefont {Shen},\ and\ \citenamefont {Forbes}}]{he2022towards}%
  \BibitemOpen
  \bibfield  {author} {\bibinfo {author} {\bibfnamefont {C.}~\bibnamefont
  {He}}, \bibinfo {author} {\bibfnamefont {Y.}~\bibnamefont {Shen}}, \ and\
  \bibinfo {author} {\bibfnamefont {A.}~\bibnamefont {Forbes}},\ }\href@noop {}
  {\bibfield  {journal} {\bibinfo  {journal} {Light: Science \& Applications}\
  }\textbf {\bibinfo {volume} {11}},\ \bibinfo {pages} {205} (\bibinfo {year}
  {2022})}\BibitemShut {NoStop}%
\bibitem [{\citenamefont {Chen}\ \emph {et~al.}(2018)\citenamefont {Chen},
  \citenamefont {Wan},\ and\ \citenamefont {Zhan}}]{chen2018vectorial}%
  \BibitemOpen
  \bibfield  {author} {\bibinfo {author} {\bibfnamefont {J.}~\bibnamefont
  {Chen}}, \bibinfo {author} {\bibfnamefont {C.}~\bibnamefont {Wan}}, \ and\
  \bibinfo {author} {\bibfnamefont {Q.}~\bibnamefont {Zhan}},\ }\href@noop {}
  {\bibfield  {journal} {\bibinfo  {journal} {Science Bulletin}\ }\textbf
  {\bibinfo {volume} {63}},\ \bibinfo {pages} {54} (\bibinfo {year}
  {2018})}\BibitemShut {NoStop}%
\bibitem [{\citenamefont {Dorn}\ \emph {et~al.}(2003)\citenamefont {Dorn},
  \citenamefont {Quabis},\ and\ \citenamefont {Leuchs}}]{dorn2003sharper}%
  \BibitemOpen
  \bibfield  {author} {\bibinfo {author} {\bibfnamefont {R.}~\bibnamefont
  {Dorn}}, \bibinfo {author} {\bibfnamefont {S.}~\bibnamefont {Quabis}}, \ and\
  \bibinfo {author} {\bibfnamefont {G.}~\bibnamefont {Leuchs}},\ }\href@noop {}
  {\bibfield  {journal} {\bibinfo  {journal} {Physical review letters}\
  }\textbf {\bibinfo {volume} {91}},\ \bibinfo {pages} {233901} (\bibinfo
  {year} {2003})}\BibitemShut {NoStop}%
\bibitem [{\citenamefont {Fickler}\ \emph {et~al.}(2014)\citenamefont
  {Fickler}, \citenamefont {Lapkiewicz}, \citenamefont {Ramelow},\ and\
  \citenamefont {Zeilinger}}]{fickler2014quantum}%
  \BibitemOpen
  \bibfield  {author} {\bibinfo {author} {\bibfnamefont {R.}~\bibnamefont
  {Fickler}}, \bibinfo {author} {\bibfnamefont {R.}~\bibnamefont {Lapkiewicz}},
  \bibinfo {author} {\bibfnamefont {S.}~\bibnamefont {Ramelow}}, \ and\
  \bibinfo {author} {\bibfnamefont {A.}~\bibnamefont {Zeilinger}},\ }\href@noop
  {} {\bibfield  {journal} {\bibinfo  {journal} {Physical Review A}\ }\textbf
  {\bibinfo {volume} {89}},\ \bibinfo {pages} {060301} (\bibinfo {year}
  {2014})}\BibitemShut {NoStop}%
\bibitem [{\citenamefont {Berg-Johansen}\ \emph {et~al.}(2015)\citenamefont
  {Berg-Johansen}, \citenamefont {T{\"o}ppel}, \citenamefont {Stiller},
  \citenamefont {Banzer}, \citenamefont {Ornigotti}, \citenamefont {Giacobino},
  \citenamefont {Leuchs}, \citenamefont {Aiello},\ and\ \citenamefont
  {Marquardt}}]{berg2015classically}%
  \BibitemOpen
  \bibfield  {author} {\bibinfo {author} {\bibfnamefont {S.}~\bibnamefont
  {Berg-Johansen}}, \bibinfo {author} {\bibfnamefont {F.}~\bibnamefont
  {T{\"o}ppel}}, \bibinfo {author} {\bibfnamefont {B.}~\bibnamefont {Stiller}},
  \bibinfo {author} {\bibfnamefont {P.}~\bibnamefont {Banzer}}, \bibinfo
  {author} {\bibfnamefont {M.}~\bibnamefont {Ornigotti}}, \bibinfo {author}
  {\bibfnamefont {E.}~\bibnamefont {Giacobino}}, \bibinfo {author}
  {\bibfnamefont {G.}~\bibnamefont {Leuchs}}, \bibinfo {author} {\bibfnamefont
  {A.}~\bibnamefont {Aiello}}, \ and\ \bibinfo {author} {\bibfnamefont
  {C.}~\bibnamefont {Marquardt}},\ }\href@noop {} {\bibfield  {journal}
  {\bibinfo  {journal} {Optica}\ }\textbf {\bibinfo {volume} {2}},\ \bibinfo
  {pages} {864} (\bibinfo {year} {2015})}\BibitemShut {NoStop}%
\bibitem [{\citenamefont {Larocque}\ \emph {et~al.}(2018)\citenamefont
  {Larocque}, \citenamefont {Sugic}, \citenamefont {Mortimer}, \citenamefont
  {Taylor}, \citenamefont {Fickler}, \citenamefont {Boyd}, \citenamefont
  {Dennis},\ and\ \citenamefont {Karimi}}]{larocque2018reconstructing}%
  \BibitemOpen
  \bibfield  {author} {\bibinfo {author} {\bibfnamefont {H.}~\bibnamefont
  {Larocque}}, \bibinfo {author} {\bibfnamefont {D.}~\bibnamefont {Sugic}},
  \bibinfo {author} {\bibfnamefont {D.}~\bibnamefont {Mortimer}}, \bibinfo
  {author} {\bibfnamefont {A.~J.}\ \bibnamefont {Taylor}}, \bibinfo {author}
  {\bibfnamefont {R.}~\bibnamefont {Fickler}}, \bibinfo {author} {\bibfnamefont
  {R.~W.}\ \bibnamefont {Boyd}}, \bibinfo {author} {\bibfnamefont {M.~R.}\
  \bibnamefont {Dennis}}, \ and\ \bibinfo {author} {\bibfnamefont
  {E.}~\bibnamefont {Karimi}},\ }\href@noop {} {\bibfield  {journal} {\bibinfo
  {journal} {Nature Physics}\ }\textbf {\bibinfo {volume} {14}},\ \bibinfo
  {pages} {1079} (\bibinfo {year} {2018})}\BibitemShut {NoStop}%
\bibitem [{\citenamefont {Milione}\ \emph {et~al.}(2011)\citenamefont
  {Milione}, \citenamefont {Sztul}, \citenamefont {Nolan},\ and\ \citenamefont
  {Alfano}}]{Milione2011}%
  \BibitemOpen
  \bibfield  {author} {\bibinfo {author} {\bibfnamefont {G.}~\bibnamefont
  {Milione}}, \bibinfo {author} {\bibfnamefont {H.~I.}\ \bibnamefont {Sztul}},
  \bibinfo {author} {\bibfnamefont {D.~A.}\ \bibnamefont {Nolan}}, \ and\
  \bibinfo {author} {\bibfnamefont {R.~R.}\ \bibnamefont {Alfano}},\
  }\href@noop {} {\bibfield  {journal} {\bibinfo  {journal} {Phys. Rev. Lett.}\
  }\textbf {\bibinfo {volume} {107}},\ \bibinfo {pages} {053601} (\bibinfo
  {year} {2011})}\BibitemShut {NoStop}%
\bibitem [{\citenamefont {Holleczek}\ \emph {et~al.}(2010)\citenamefont
  {Holleczek}, \citenamefont {Aiello}, \citenamefont {Gabriel}, \citenamefont
  {Marquardt},\ and\ \citenamefont {Leuchs}}]{holleczek2010poincare}%
  \BibitemOpen
  \bibfield  {author} {\bibinfo {author} {\bibfnamefont {A.}~\bibnamefont
  {Holleczek}}, \bibinfo {author} {\bibfnamefont {A.}~\bibnamefont {Aiello}},
  \bibinfo {author} {\bibfnamefont {C.}~\bibnamefont {Gabriel}}, \bibinfo
  {author} {\bibfnamefont {C.}~\bibnamefont {Marquardt}}, \ and\ \bibinfo
  {author} {\bibfnamefont {G.}~\bibnamefont {Leuchs}},\ }\href@noop {}
  {\bibfield  {journal} {\bibinfo  {journal} {arXiv preprint arXiv:1007.2528}\
  } (\bibinfo {year} {2010})}\BibitemShut {NoStop}%
\bibitem [{\citenamefont {Milione}\ \emph {et~al.}(2012)\citenamefont
  {Milione}, \citenamefont {Evans}, \citenamefont {Nolan},\ and\ \citenamefont
  {Alfano}}]{milione2012higher}%
  \BibitemOpen
  \bibfield  {author} {\bibinfo {author} {\bibfnamefont {G.}~\bibnamefont
  {Milione}}, \bibinfo {author} {\bibfnamefont {S.}~\bibnamefont {Evans}},
  \bibinfo {author} {\bibfnamefont {D.}~\bibnamefont {Nolan}}, \ and\ \bibinfo
  {author} {\bibfnamefont {R.}~\bibnamefont {Alfano}},\ }\href@noop {}
  {\bibfield  {journal} {\bibinfo  {journal} {Physical Review Letters}\
  }\textbf {\bibinfo {volume} {108}},\ \bibinfo {pages} {190401} (\bibinfo
  {year} {2012})}\BibitemShut {NoStop}%
\bibitem [{\citenamefont {Ndagano}\ \emph {et~al.}(2017)\citenamefont
  {Ndagano}, \citenamefont {Perez-Garcia}, \citenamefont {Roux}, \citenamefont
  {McLaren}, \citenamefont {Rosales-Guzman}, \citenamefont {Zhang},
  \citenamefont {Mouane}, \citenamefont {Hernandez-Aranda}, \citenamefont
  {Konrad},\ and\ \citenamefont {Forbes}}]{ndagano2017characterizing}%
  \BibitemOpen
  \bibfield  {author} {\bibinfo {author} {\bibfnamefont {B.}~\bibnamefont
  {Ndagano}}, \bibinfo {author} {\bibfnamefont {B.}~\bibnamefont
  {Perez-Garcia}}, \bibinfo {author} {\bibfnamefont {F.~S.}\ \bibnamefont
  {Roux}}, \bibinfo {author} {\bibfnamefont {M.}~\bibnamefont {McLaren}},
  \bibinfo {author} {\bibfnamefont {C.}~\bibnamefont {Rosales-Guzman}},
  \bibinfo {author} {\bibfnamefont {Y.}~\bibnamefont {Zhang}}, \bibinfo
  {author} {\bibfnamefont {O.}~\bibnamefont {Mouane}}, \bibinfo {author}
  {\bibfnamefont {R.~I.}\ \bibnamefont {Hernandez-Aranda}}, \bibinfo {author}
  {\bibfnamefont {T.}~\bibnamefont {Konrad}}, \ and\ \bibinfo {author}
  {\bibfnamefont {A.}~\bibnamefont {Forbes}},\ }\href@noop {} {\bibfield
  {journal} {\bibinfo  {journal} {Nature Physics}\ }\textbf {\bibinfo {volume}
  {13}},\ \bibinfo {pages} {397} (\bibinfo {year} {2017})}\BibitemShut
  {NoStop}%
\bibitem [{\citenamefont {Jiang}\ \emph {et~al.}(2020)\citenamefont {Jiang},
  \citenamefont {Kang}, \citenamefont {Yue}, \citenamefont {Xu}, \citenamefont
  {Yang}, \citenamefont {Jin}, \citenamefont {Yu}, \citenamefont {Hong},
  \citenamefont {Werner},\ and\ \citenamefont {Qiu}}]{jiang2020single}%
  \BibitemOpen
  \bibfield  {author} {\bibinfo {author} {\bibfnamefont {Z.~H.}\ \bibnamefont
  {Jiang}}, \bibinfo {author} {\bibfnamefont {L.}~\bibnamefont {Kang}},
  \bibinfo {author} {\bibfnamefont {T.}~\bibnamefont {Yue}}, \bibinfo {author}
  {\bibfnamefont {H.-X.}\ \bibnamefont {Xu}}, \bibinfo {author} {\bibfnamefont
  {Y.}~\bibnamefont {Yang}}, \bibinfo {author} {\bibfnamefont {Z.}~\bibnamefont
  {Jin}}, \bibinfo {author} {\bibfnamefont {C.}~\bibnamefont {Yu}}, \bibinfo
  {author} {\bibfnamefont {W.}~\bibnamefont {Hong}}, \bibinfo {author}
  {\bibfnamefont {D.~H.}\ \bibnamefont {Werner}}, \ and\ \bibinfo {author}
  {\bibfnamefont {C.-W.}\ \bibnamefont {Qiu}},\ }\href@noop {} {\bibfield
  {journal} {\bibinfo  {journal} {Advanced Materials}\ }\textbf {\bibinfo
  {volume} {32}},\ \bibinfo {pages} {1903983} (\bibinfo {year}
  {2020})}\BibitemShut {NoStop}%
\bibitem [{\citenamefont {Shen}\ \emph {et~al.}(2020)\citenamefont {Shen},
  \citenamefont {Wang}, \citenamefont {Fu}, \citenamefont {Naidoo},\ and\
  \citenamefont {Forbes}}]{shen20202}%
  \BibitemOpen
  \bibfield  {author} {\bibinfo {author} {\bibfnamefont {Y.}~\bibnamefont
  {Shen}}, \bibinfo {author} {\bibfnamefont {Z.}~\bibnamefont {Wang}}, \bibinfo
  {author} {\bibfnamefont {X.}~\bibnamefont {Fu}}, \bibinfo {author}
  {\bibfnamefont {D.}~\bibnamefont {Naidoo}}, \ and\ \bibinfo {author}
  {\bibfnamefont {A.}~\bibnamefont {Forbes}},\ }\href@noop {} {\bibfield
  {journal} {\bibinfo  {journal} {Physical Review A}\ }\textbf {\bibinfo
  {volume} {102}},\ \bibinfo {pages} {031501} (\bibinfo {year}
  {2020})}\BibitemShut {NoStop}%
\bibitem [{\citenamefont {Beckley}\ \emph
  {et~al.}(2010{\natexlab{a}})\citenamefont {Beckley}, \citenamefont {Brown},\
  and\ \citenamefont {Alonso}}]{beckley2010full}%
  \BibitemOpen
  \bibfield  {author} {\bibinfo {author} {\bibfnamefont {A.~M.}\ \bibnamefont
  {Beckley}}, \bibinfo {author} {\bibfnamefont {T.~G.}\ \bibnamefont {Brown}},
  \ and\ \bibinfo {author} {\bibfnamefont {M.~A.}\ \bibnamefont {Alonso}},\
  }\href@noop {} {\bibfield  {journal} {\bibinfo  {journal} {Optics express}\
  }\textbf {\bibinfo {volume} {18}},\ \bibinfo {pages} {10777} (\bibinfo {year}
  {2010}{\natexlab{a}})}\BibitemShut {NoStop}%
\bibitem [{\citenamefont {Dennis}(2002)}]{dennis2002polarization}%
  \BibitemOpen
  \bibfield  {author} {\bibinfo {author} {\bibfnamefont {M.}~\bibnamefont
  {Dennis}},\ }\href@noop {} {\bibfield  {journal} {\bibinfo  {journal} {Optics
  Communications}\ }\textbf {\bibinfo {volume} {213}},\ \bibinfo {pages} {201}
  (\bibinfo {year} {2002})}\BibitemShut {NoStop}%
\bibitem [{\citenamefont {Bauer}\ \emph {et~al.}(2015)\citenamefont {Bauer},
  \citenamefont {Banzer}, \citenamefont {Karimi}, \citenamefont {Orlov},
  \citenamefont {Rubano}, \citenamefont {Marrucci}, \citenamefont {Santamato},
  \citenamefont {Boyd},\ and\ \citenamefont {Leuchs}}]{bauer2015observation}%
  \BibitemOpen
  \bibfield  {author} {\bibinfo {author} {\bibfnamefont {T.}~\bibnamefont
  {Bauer}}, \bibinfo {author} {\bibfnamefont {P.}~\bibnamefont {Banzer}},
  \bibinfo {author} {\bibfnamefont {E.}~\bibnamefont {Karimi}}, \bibinfo
  {author} {\bibfnamefont {S.}~\bibnamefont {Orlov}}, \bibinfo {author}
  {\bibfnamefont {A.}~\bibnamefont {Rubano}}, \bibinfo {author} {\bibfnamefont
  {L.}~\bibnamefont {Marrucci}}, \bibinfo {author} {\bibfnamefont
  {E.}~\bibnamefont {Santamato}}, \bibinfo {author} {\bibfnamefont {R.~W.}\
  \bibnamefont {Boyd}}, \ and\ \bibinfo {author} {\bibfnamefont
  {G.}~\bibnamefont {Leuchs}},\ }\href@noop {} {\bibfield  {journal} {\bibinfo
  {journal} {Science}\ }\textbf {\bibinfo {volume} {347}},\ \bibinfo {pages}
  {964} (\bibinfo {year} {2015})}\BibitemShut {NoStop}%
\bibitem [{\citenamefont {Shen}\ \emph {et~al.}(2024)\citenamefont {Shen},
  \citenamefont {Zhang}, \citenamefont {Shi}, \citenamefont {Du}, \citenamefont
  {Yuan},\ and\ \citenamefont {Zayats}}]{shen2024optical}%
  \BibitemOpen
  \bibfield  {author} {\bibinfo {author} {\bibfnamefont {Y.}~\bibnamefont
  {Shen}}, \bibinfo {author} {\bibfnamefont {Q.}~\bibnamefont {Zhang}},
  \bibinfo {author} {\bibfnamefont {P.}~\bibnamefont {Shi}}, \bibinfo {author}
  {\bibfnamefont {L.}~\bibnamefont {Du}}, \bibinfo {author} {\bibfnamefont
  {X.}~\bibnamefont {Yuan}}, \ and\ \bibinfo {author} {\bibfnamefont {A.~V.}\
  \bibnamefont {Zayats}},\ }\href@noop {} {\bibfield  {journal} {\bibinfo
  {journal} {Nature Photonics}\ }\textbf {\bibinfo {volume} {18}},\ \bibinfo
  {pages} {15} (\bibinfo {year} {2024})}\BibitemShut {NoStop}%
\bibitem [{\citenamefont {Ornelas}\ \emph {et~al.}(2024)\citenamefont
  {Ornelas}, \citenamefont {Nape}, \citenamefont {de~Mello~Koch},\ and\
  \citenamefont {Forbes}}]{ornelas2024non}%
  \BibitemOpen
  \bibfield  {author} {\bibinfo {author} {\bibfnamefont {P.}~\bibnamefont
  {Ornelas}}, \bibinfo {author} {\bibfnamefont {I.}~\bibnamefont {Nape}},
  \bibinfo {author} {\bibfnamefont {R.}~\bibnamefont {de~Mello~Koch}}, \ and\
  \bibinfo {author} {\bibfnamefont {A.}~\bibnamefont {Forbes}},\ }\href@noop {}
  {\bibfield  {journal} {\bibinfo  {journal} {Nature Photonics}\ ,\ \bibinfo
  {pages} {1}} (\bibinfo {year} {2024})}\BibitemShut {NoStop}%
\bibitem [{\citenamefont {Padgett}\ and\ \citenamefont
  {Courtial}(1999)}]{padgett1999poincare}%
  \BibitemOpen
  \bibfield  {author} {\bibinfo {author} {\bibfnamefont {M.~J.}\ \bibnamefont
  {Padgett}}\ and\ \bibinfo {author} {\bibfnamefont {J.}~\bibnamefont
  {Courtial}},\ }\href@noop {} {\bibfield  {journal} {\bibinfo  {journal}
  {Optics letters}\ }\textbf {\bibinfo {volume} {24}},\ \bibinfo {pages} {430}
  (\bibinfo {year} {1999})}\BibitemShut {NoStop}%
\bibitem [{\citenamefont {Agarwal}(1999)}]{agarwal19992}%
  \BibitemOpen
  \bibfield  {author} {\bibinfo {author} {\bibfnamefont {G.}~\bibnamefont
  {Agarwal}},\ }\href@noop {} {\bibfield  {journal} {\bibinfo  {journal} {JOSA
  A}\ }\textbf {\bibinfo {volume} {16}},\ \bibinfo {pages} {2914} (\bibinfo
  {year} {1999})}\BibitemShut {NoStop}%
\bibitem [{\citenamefont {Dennis}\ and\ \citenamefont
  {Alonso}(2017)}]{dennis2017swings}%
  \BibitemOpen
  \bibfield  {author} {\bibinfo {author} {\bibfnamefont {M.~R.}\ \bibnamefont
  {Dennis}}\ and\ \bibinfo {author} {\bibfnamefont {M.~A.}\ \bibnamefont
  {Alonso}},\ }\href@noop {} {\bibfield  {journal} {\bibinfo  {journal}
  {Philosophical Transactions of the Royal Society A: Mathematical, Physical
  and Engineering Sciences}\ }\textbf {\bibinfo {volume} {375}},\ \bibinfo
  {pages} {20150441} (\bibinfo {year} {2017})}\BibitemShut {NoStop}%
\bibitem [{\citenamefont {Brecht}\ \emph {et~al.}(2015)\citenamefont {Brecht},
  \citenamefont {Reddy}, \citenamefont {Silberhorn},\ and\ \citenamefont
  {Raymer}}]{brecht2015photon}%
  \BibitemOpen
  \bibfield  {author} {\bibinfo {author} {\bibfnamefont {B.}~\bibnamefont
  {Brecht}}, \bibinfo {author} {\bibfnamefont {D.~V.}\ \bibnamefont {Reddy}},
  \bibinfo {author} {\bibfnamefont {C.}~\bibnamefont {Silberhorn}}, \ and\
  \bibinfo {author} {\bibfnamefont {M.~G.}\ \bibnamefont {Raymer}},\
  }\href@noop {} {\bibfield  {journal} {\bibinfo  {journal} {Physical Review
  X}\ }\textbf {\bibinfo {volume} {5}},\ \bibinfo {pages} {041017} (\bibinfo
  {year} {2015})}\BibitemShut {NoStop}%
\bibitem [{\citenamefont {Nielsen}\ and\ \citenamefont
  {Chuang}(2010)}]{nielsen2010quantum}%
  \BibitemOpen
  \bibfield  {author} {\bibinfo {author} {\bibfnamefont {M.~A.}\ \bibnamefont
  {Nielsen}}\ and\ \bibinfo {author} {\bibfnamefont {I.~L.}\ \bibnamefont
  {Chuang}},\ }\href@noop {} {\emph {\bibinfo {title} {Quantum computation and
  quantum information}}}\ (\bibinfo  {publisher} {Cambridge university press},\
  \bibinfo {year} {2010})\BibitemShut {NoStop}%
\bibitem [{\citenamefont {Kopf}\ \emph {et~al.}(2021)\citenamefont {Kopf},
  \citenamefont {Ruano}, \citenamefont {Hiekkam{\"a}ki}, \citenamefont {Stolt},
  \citenamefont {Huttunen}, \citenamefont {Bouchard},\ and\ \citenamefont
  {Fickler}}]{kopf2021spectral}%
  \BibitemOpen
  \bibfield  {author} {\bibinfo {author} {\bibfnamefont {L.}~\bibnamefont
  {Kopf}}, \bibinfo {author} {\bibfnamefont {J.~R.~D.}\ \bibnamefont {Ruano}},
  \bibinfo {author} {\bibfnamefont {M.}~\bibnamefont {Hiekkam{\"a}ki}},
  \bibinfo {author} {\bibfnamefont {T.}~\bibnamefont {Stolt}}, \bibinfo
  {author} {\bibfnamefont {M.~J.}\ \bibnamefont {Huttunen}}, \bibinfo {author}
  {\bibfnamefont {F.}~\bibnamefont {Bouchard}}, \ and\ \bibinfo {author}
  {\bibfnamefont {R.}~\bibnamefont {Fickler}},\ }\href@noop {} {\bibfield
  {journal} {\bibinfo  {journal} {Optica}\ }\textbf {\bibinfo {volume} {8}},\
  \bibinfo {pages} {930} (\bibinfo {year} {2021})}\BibitemShut {NoStop}%
\bibitem [{\citenamefont {Kopf}\ \emph {et~al.}(2023)\citenamefont {Kopf},
  \citenamefont {Barros},\ and\ \citenamefont {Fickler}}]{kopf2023correlating}%
  \BibitemOpen
  \bibfield  {author} {\bibinfo {author} {\bibfnamefont {L.}~\bibnamefont
  {Kopf}}, \bibinfo {author} {\bibfnamefont {R.}~\bibnamefont {Barros}}, \ and\
  \bibinfo {author} {\bibfnamefont {R.}~\bibnamefont {Fickler}},\ }\href@noop
  {} {\bibfield  {journal} {\bibinfo  {journal} {ACS Photonics}\ }\textbf
  {\bibinfo {volume} {11}},\ \bibinfo {pages} {241} (\bibinfo {year}
  {2023})}\BibitemShut {NoStop}%
\bibitem [{\citenamefont {Bolduc}\ \emph {et~al.}(2013)\citenamefont {Bolduc},
  \citenamefont {Bent}, \citenamefont {Santamato}, \citenamefont {Karimi},\
  and\ \citenamefont {Boyd}}]{bolduc2013exact}%
  \BibitemOpen
  \bibfield  {author} {\bibinfo {author} {\bibfnamefont {E.}~\bibnamefont
  {Bolduc}}, \bibinfo {author} {\bibfnamefont {N.}~\bibnamefont {Bent}},
  \bibinfo {author} {\bibfnamefont {E.}~\bibnamefont {Santamato}}, \bibinfo
  {author} {\bibfnamefont {E.}~\bibnamefont {Karimi}}, \ and\ \bibinfo {author}
  {\bibfnamefont {R.~W.}\ \bibnamefont {Boyd}},\ }\href@noop {} {\bibfield
  {journal} {\bibinfo  {journal} {Optics letters}\ }\textbf {\bibinfo {volume}
  {38}},\ \bibinfo {pages} {3546} (\bibinfo {year} {2013})}\BibitemShut
  {NoStop}%
\bibitem [{\citenamefont {Weiner}(2011)}]{weiner2011ultrafast}%
  \BibitemOpen
  \bibfield  {author} {\bibinfo {author} {\bibfnamefont {A.~M.}\ \bibnamefont
  {Weiner}},\ }\href@noop {} {\bibfield  {journal} {\bibinfo  {journal} {Optics
  Communications}\ }\textbf {\bibinfo {volume} {284}},\ \bibinfo {pages} {3669}
  (\bibinfo {year} {2011})}\BibitemShut {NoStop}%
\bibitem [{\citenamefont {Biener}\ \emph {et~al.}(2002)\citenamefont {Biener},
  \citenamefont {Niv}, \citenamefont {Kleiner},\ and\ \citenamefont
  {Hasman}}]{biener2002formation}%
  \BibitemOpen
  \bibfield  {author} {\bibinfo {author} {\bibfnamefont {G.}~\bibnamefont
  {Biener}}, \bibinfo {author} {\bibfnamefont {A.}~\bibnamefont {Niv}},
  \bibinfo {author} {\bibfnamefont {V.}~\bibnamefont {Kleiner}}, \ and\
  \bibinfo {author} {\bibfnamefont {E.}~\bibnamefont {Hasman}},\ }\href@noop {}
  {\bibfield  {journal} {\bibinfo  {journal} {Optics letters}\ }\textbf
  {\bibinfo {volume} {27}},\ \bibinfo {pages} {1875} (\bibinfo {year}
  {2002})}\BibitemShut {NoStop}%
\bibitem [{\citenamefont {Pisanty}\ \emph {et~al.}(2019)\citenamefont
  {Pisanty}, \citenamefont {Machado}, \citenamefont {Vicu{\~n}a-Hern{\'a}ndez},
  \citenamefont {Pic{\'o}n}, \citenamefont {Celi}, \citenamefont {Torres},\
  and\ \citenamefont {Lewenstein}}]{pisanty2019knotting}%
  \BibitemOpen
  \bibfield  {author} {\bibinfo {author} {\bibfnamefont {E.}~\bibnamefont
  {Pisanty}}, \bibinfo {author} {\bibfnamefont {G.~J.}\ \bibnamefont
  {Machado}}, \bibinfo {author} {\bibfnamefont {V.}~\bibnamefont
  {Vicu{\~n}a-Hern{\'a}ndez}}, \bibinfo {author} {\bibfnamefont
  {A.}~\bibnamefont {Pic{\'o}n}}, \bibinfo {author} {\bibfnamefont
  {A.}~\bibnamefont {Celi}}, \bibinfo {author} {\bibfnamefont {J.~P.}\
  \bibnamefont {Torres}}, \ and\ \bibinfo {author} {\bibfnamefont
  {M.}~\bibnamefont {Lewenstein}},\ }\href@noop {} {\bibfield  {journal}
  {\bibinfo  {journal} {Nature Photonics}\ }\textbf {\bibinfo {volume} {13}},\
  \bibinfo {pages} {569} (\bibinfo {year} {2019})}\BibitemShut {NoStop}%
\bibitem [{\citenamefont {Sugic}\ \emph {et~al.}(2020)\citenamefont {Sugic},
  \citenamefont {Dennis}, \citenamefont {Nori},\ and\ \citenamefont
  {Bliokh}}]{sugic2020knotted}%
  \BibitemOpen
  \bibfield  {author} {\bibinfo {author} {\bibfnamefont {D.}~\bibnamefont
  {Sugic}}, \bibinfo {author} {\bibfnamefont {M.~R.}\ \bibnamefont {Dennis}},
  \bibinfo {author} {\bibfnamefont {F.}~\bibnamefont {Nori}}, \ and\ \bibinfo
  {author} {\bibfnamefont {K.~Y.}\ \bibnamefont {Bliokh}},\ }\href@noop {}
  {\bibfield  {journal} {\bibinfo  {journal} {Physical Review Research}\
  }\textbf {\bibinfo {volume} {2}},\ \bibinfo {pages} {042045} (\bibinfo {year}
  {2020})}\BibitemShut {NoStop}%
\bibitem [{\citenamefont {Lévay}(2004)}]{fibration1}%
  \BibitemOpen
  \bibfield  {author} {\bibinfo {author} {\bibfnamefont {P.}~\bibnamefont
  {Lévay}},\ }\href@noop {} {\bibfield  {journal} {\bibinfo  {journal} {J.
  Phys. A: Math. Gen.}\ }\textbf {\bibinfo {volume} {37}},\ \bibinfo {pages}
  {1821} (\bibinfo {year} {2004})}\BibitemShut {NoStop}%
\bibitem [{\citenamefont {Bengtsson}\ and\ \citenamefont
  {Życkowski}(2020)}]{fibration2}%
  \BibitemOpen
  \bibfield  {author} {\bibinfo {author} {\bibfnamefont {I.}~\bibnamefont
  {Bengtsson}}\ and\ \bibinfo {author} {\bibfnamefont {K.}~\bibnamefont
  {Życkowski}},\ }\href@noop {} {\emph {\bibinfo {title} {Geometry of quantum
  states - An introduction to quantum entanglement}}}\ (\bibinfo  {publisher}
  {Cambridge university press},\ \bibinfo {year} {2020})\BibitemShut {NoStop}%
\bibitem [{\citenamefont {Peters}\ \emph {et~al.}(2023)\citenamefont {Peters},
  \citenamefont {Ornelas}, \citenamefont {Nape},\ and\ \citenamefont
  {Forbes}}]{peters2023spatially}%
  \BibitemOpen
  \bibfield  {author} {\bibinfo {author} {\bibfnamefont {C.}~\bibnamefont
  {Peters}}, \bibinfo {author} {\bibfnamefont {P.}~\bibnamefont {Ornelas}},
  \bibinfo {author} {\bibfnamefont {I.}~\bibnamefont {Nape}}, \ and\ \bibinfo
  {author} {\bibfnamefont {A.}~\bibnamefont {Forbes}},\ }\href@noop {}
  {\bibfield  {journal} {\bibinfo  {journal} {Physical Review A}\ }\textbf
  {\bibinfo {volume} {108}},\ \bibinfo {pages} {053502} (\bibinfo {year}
  {2023})}\BibitemShut {NoStop}%
\bibitem [{\citenamefont {Andrews}\ and\ \citenamefont
  {(editors)}(2013)}]{andrews}%
  \BibitemOpen
  \bibfield  {author} {\bibinfo {author} {\bibfnamefont {D.}~\bibnamefont
  {Andrews}}\ and\ \bibinfo {author} {\bibfnamefont {M.~B.}\ \bibnamefont
  {(editors)}},\ }\href@noop {} {\emph {\bibinfo {title} {The Orbital Angular
  Momentum of Light}}}\ (\bibinfo  {publisher} {Cambridge university Press},\
  \bibinfo {year} {2013})\BibitemShut {NoStop}%
\bibitem [{\citenamefont {Brixner}\ and\ \citenamefont
  {Gerber}(2001)}]{brixner2001femtosecond}%
  \BibitemOpen
  \bibfield  {author} {\bibinfo {author} {\bibfnamefont {T.}~\bibnamefont
  {Brixner}}\ and\ \bibinfo {author} {\bibfnamefont {G.}~\bibnamefont
  {Gerber}},\ }\href@noop {} {\bibfield  {journal} {\bibinfo  {journal} {Optics
  letters}\ }\textbf {\bibinfo {volume} {26}},\ \bibinfo {pages} {557}
  (\bibinfo {year} {2001})}\BibitemShut {NoStop}%
\bibitem [{\citenamefont {Beckley}\ \emph
  {et~al.}(2010{\natexlab{b}})\citenamefont {Beckley}, \citenamefont
  {Thomas~G.},\ and\ \citenamefont {Miguel~A.}}]{alonsoFullPB}%
  \BibitemOpen
  \bibfield  {author} {\bibinfo {author} {\bibfnamefont {A.~M.}\ \bibnamefont
  {Beckley}}, \bibinfo {author} {\bibfnamefont {B.}~\bibnamefont {Thomas~G.}},
  \ and\ \bibinfo {author} {\bibfnamefont {A.}~\bibnamefont {Miguel~A.}},\
  }\href@noop {} {\bibfield  {journal} {\bibinfo  {journal} {Opt. Express}\
  }\textbf {\bibinfo {volume} {18}},\ \bibinfo {pages} {10777} (\bibinfo {year}
  {2010}{\natexlab{b}})}\BibitemShut {NoStop}%
\end{thebibliography}%

\onecolumngrid

\section{Supplementary}
This Supplementary material is structured as follows: in Sec.~\ref{sec:1} we briefly recall the conventions and definition that allow one to represent an arbitrary polarization state on the Poincaré sphere (PS) and derive the expression for the Stokes parameters of an arbitrarily polarised beam within the framework of the PS representation. Sec.~\ref{sec:2} and \ref{sec:3} are  dedicated to briefly recap the extension of the concept of the PS to spatial and spectral modes, respectively, and the subsequent definition of the higher-order PS. These concepts are then combined in Sec.~\ref{sec:4} to introduce the spatio-spectral PS and calculate the Stokes parameters within this formalism. Finally, Sec.~\ref{sec:5} gives more details about the experimental generation of spatio-spectral beams.

\subsection{Poincaré sphere}
\label{sec:1}
\noindent In this section, we introduce the convention used in the manuscript for defining the PS. Any polarization state of light can be described as a two-dimensional state using the circular polarisation states $\ket{R}$ and $\ket{L}$, i.e.,
\begin{equation}\label{eqA:pol_state}
\ket{\Psi}=\cos\left(\frac{\theta}{2}\right)\ket{R} + \sin\left(\frac{\theta}{2}\right) e^{i\phi} \ket{L}.  
\end{equation}
Here, $\theta\in[0,\pi]$ and $\phi\in[0,2\pi]$, and the orthonormal circular polarization basis is defined as 
\begin{eqnarray}
\ket{L}&=&\frac{1}{\sqrt{2}}\left(\Vec{x}+i\Vec{y}\right),\\
\ket{R}&=&\frac{1}{\sqrt{2}}\left(\Vec{x}-i\Vec{y}\right),
\end{eqnarray}
where $\Vec{x}$ and $\Vec{y}$ are the transverse Cartesian coordinates of the light field. Notice, that with this choice of parametrization, $\ket{R}$ is located on the north-pole of the PS, while $\ket{L}$ is on the south-pole. The equator of the sphere is spanned by the linear polarization states
\begin{eqnarray}
\ket{H}&=& \frac{1}{\sqrt{2}}\left(\ket{R} + \ket{L}\right) = \Vec{x}, \\
\ket{V}&=& \frac{1}{\sqrt{2}}\left(\ket{R} - \ket{L}\right) = \Vec{y}, \\
\ket{D}&=& \frac{1}{\sqrt{2}}\left(\ket{R} -i \ket{L}\right) = \frac{1}{\sqrt{2}}\left(\Vec{x}+\Vec{y}\right),\\
\ket{A}&=& \frac{1}{\sqrt{2}}\left(\ket{R} +i \ket{L}\right) = \frac{1}{\sqrt{2}}\left(\Vec{x}-\Vec{y}\right),
\end{eqnarray}
corresponding to horizontal (H), vertical (V), diagonal (D) and anti-diagonal (A) polarisation, respectively. The Stokes parameters are then defined as usual, i.e.,
\begin{subequations}
\begin{align}
S_0&= |\braket{R}{\Psi}|^2 + |\braket{L}{\Psi}|^2 =|\braket{H}{\Psi}|^2 + |\braket{V}{\Psi}|^2=|\braket{D}{\Psi}|^2 + |\braket{A}{\Psi}|^2, \label{eq:s0}\\
S_1&= |\braket{H}{\Psi}|^2 - |\braket{V}{\Psi}|^2,\\
S_2&= |\braket{D}{\Psi}|^2 - |\braket{A}{\Psi}|^2,\\
S_3&= |\braket{R}{\Psi}|^2 - |\braket{L}{\Psi}|^2.\\
\end{align}
\end{subequations}

For a general state as described in Eq.~\eqref{eqA:pol_state}, the above definitions result in
\begin{subequations}
\begin{align}
S_0&=1,\\
S_1/S_0&=2\cos(\frac{\theta}{2})\sin(\frac{\theta}{2})\cos(\phi)=\sin(\theta)\cos(\phi),\\
S_2/S_0&=2\cos(\frac{\theta}{2})\sin(\frac{\theta}{2})\sin(\phi)=\sin(\theta)\sin(\phi),\\
S_3/S_0&=\cos^2\left(\frac{\theta}{2}\right)-\sin^2\left(\frac{\theta}{2}\right)=\cos(\theta).
\end{align}
\end{subequations}
\subsection{Higher-Order Poincaré sphere in space}
\label{sec:2}
\noindent Let us now consider not only polarisation as an accessible degree of freedom (DOF) of the field, but also its transverse spatial structure.
Let's focus in particular on the first-order modes of Laguerre-Gaussian (LG) beams, which show a twisted phase front and are known to carry orbital angular momentum (OAM) \cite{andrews}. A detailed analysis of the resulting PS, i.e., the so-called higher-order PS can be found, for example, in Refs. \cite{Milione2011} and \cite{holleczek2010poincare}. Here, we just briefly recap the relevant results for our discussion in Sec.~\ref{sec:4}.
%
%

\noindent In general, one can write a hybrid space-polarization state by adding two orthogonal transverse spatial mode functions $T_{n}(\Vec{r})$ into the expression of an arbitrarily polarised state \eqref{eqA:pol_state} such that
%
\begin{eqnarray}
\ket{\Psi}_T=\cos\left(\frac{\theta_T}{2}\right)T_{1}(\Vec{r})\ket{R} + \sin\left(\frac{\theta_T}{2}\right) e^{i\phi_T} T_{2}(\Vec{r}) \ket{L},    \label{eqA:SpaVec_state}
\end{eqnarray}
with $\Vec{r}$ being the transverse spatial coordinate and, as before, $\theta_T\in[0,\pi]$ and $\phi_T\in[0,2\pi]$.
Analogue to fundamental polarization states, $\ket{\Psi}_T$ can be considered as a state in a so-called higher-order PS, where the two poles are orthogonal polarization states, each with an orthogonal spatial mode function attached to them.
The angles $\theta_T$ and $\phi_T$ can then be understood as the angular positions on the surface of this higher-order PS.
Note, that in this picture one can define mutually unbiased basis states as well as Stokes parameters in an analogous way as described above.

$T_{n}(\Vec{r})$ can be implemented with the first-order LG modes, i.e., modes with $\{p=0,\ell=\pm 1\}$, which carry one quanta of OAM. For this particular choice, the transverse field is easiest described in cylindrical coordinates with the radial position $r=\abs{x+iy}$ and the angle $\varphi=\arg(x+iy)$ such that
\begin{eqnarray}
T_{\pm 1}(r,\varphi) = G(r) e^{\pm i\varphi},
\end{eqnarray}
with $G(\Vec{r})$ describing a transverse Gaussian profile.
By replacing $T_1$ and $T_2$ in Eq.~\eqref{eqA:SpaVec_state} with $T_{-1}$ and $T_{1}$, respectively, one obtains
\begin{eqnarray}
\ket{\Psi}_T=G(r)e^{-i\varphi}\left[\cos\left(\frac{\theta_T}{2}\right)\ket{R} + \sin\left(\frac{\theta_T}{2}\right) e^{i\phi_T}e^{i2\varphi} \ket{L}\right],    \label{eqA:SpaVec_state_LG}
\end{eqnarray}
Note that, for $T_1=T_2$ or considering a plane wave (as often done), the higher-order PS reduces to the fundamental one. 

\begin{figure}[ht] 
    \centering
    \includegraphics[width=0.8\linewidth]{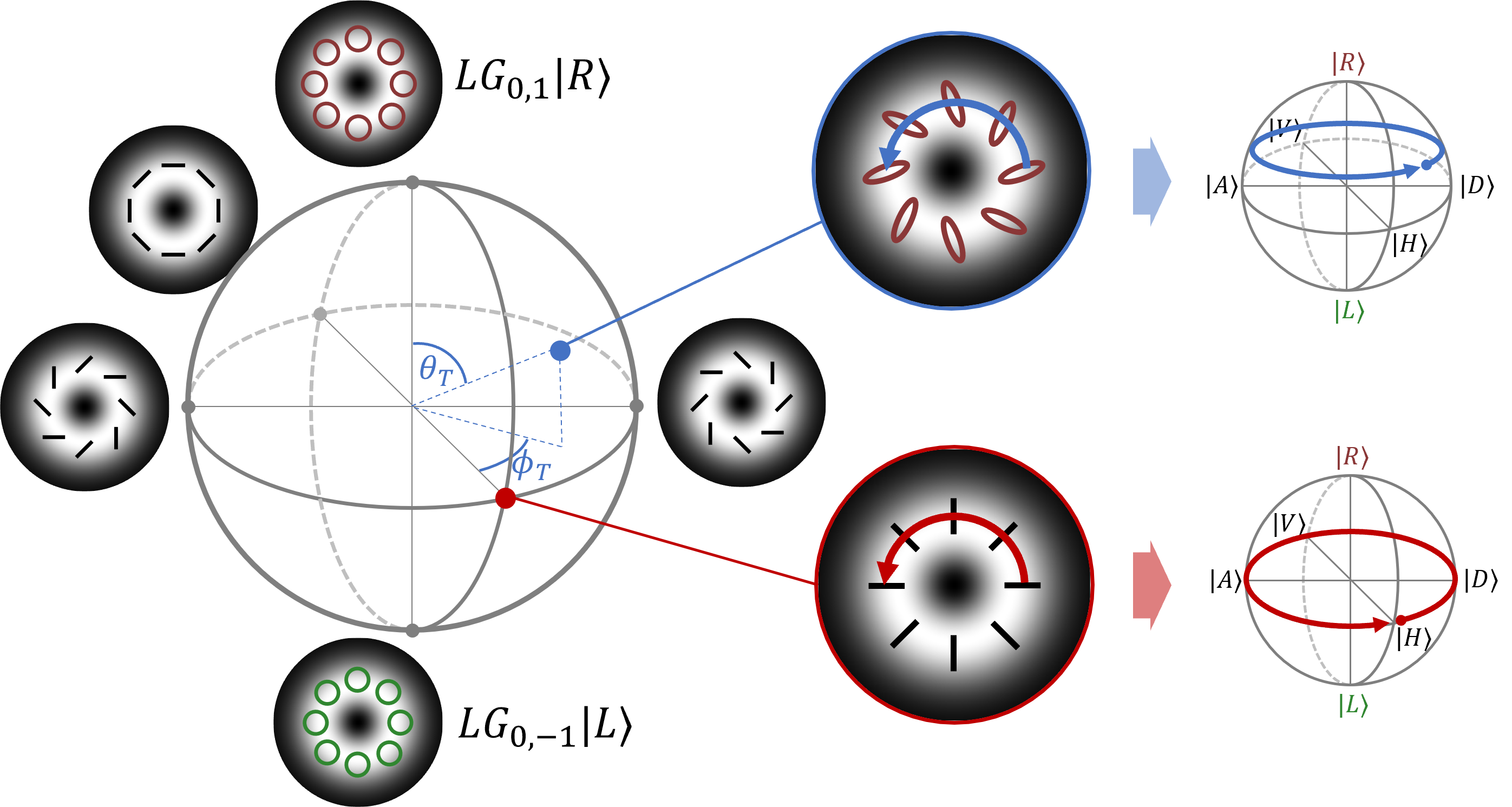}
    \caption{Higher-order Poincaré Sphere (PS) for spatial vector beams. Every point of the higher-order PS corresponds to a spatial mode for which the polarisation vector changes with the transverse angle $\varphi$, such that it follows a latitude on the fundamental PS.
    This is depicted for two examples. The upper example (highlighted in blue) is shown for $\theta_F=2\pi/3$ and $\phi_F=\pi/4$, while the lower example (highlighted in red) shows a state on the equator which corresponds to a radially polarized beam with $\theta_F=\pi/2$ and $\phi_F=0$.}
    \label{figS:0a}
\end{figure}

The Stokes parameters for a state on the higher-order PS are then given by
\begin{subequations}
\begin{align}
S_0&=|G(r)|^2,\\
S_1/S_0&=2\cos(\frac{\theta_T}{2})\sin(\frac{\theta_T}{2})\cos(\phi_T-2\varphi),\\
S_2/S_0&=2\cos(\frac{\theta_T}{2})\sin(\frac{\theta_T}{2})\sin(\phi_T-2\varphi),\\
S_3/S_0&=\cos^2\left(\frac{\theta_T}{2}\right)-\sin^2\left(\frac{\theta_T}{2}\right),
\end{align}
\end{subequations}
which can be simplified to
\begin{subequations}
\begin{align}
S_1/S_0&=\sin(\theta_T)\cos(\phi_T-2\varphi)\\
S_2/S_0&=\sin(\theta_T)\sin(\phi_T-2\varphi)\\
S_3/S_0&=\cos(\theta_T).
\end{align}
\end{subequations}
From the expression above, we see that for every point on this higher-order PS (with exception of the two poles, i.e., $\theta_T\neq0,\pi/2$), 
the two normalized Stokes parameters $S_1/S_0$ and $S_2/S_0$ are cyclically changing.
This means that for varying values of the angular coordinate $\varphi$ the polarization changes along a parallel line (horizontal circle) on the fundamental PS.
The higher-order PS for spatial vector beams is shown in Fig.\ref{figS:0a} along with two explicit examples on how the transverse varying polarization relates to the fundamental PS.

\subsection{Higher-Order Poincaré sphere in frequency (time)}
\label{sec:3}
\noindent A link similar to the one shown above can be made for the  polarisation and mode functions by combining the polarization DoF with the spectral or temporal DoF of light. 
Note, that in the following this novel type of a higher-order PS is described by linking spectral modes to polarization and thereby constructing so-called spectral vector beams \cite{kopf2021spectral}.
The same argument holds for linking temporal modes to polarization \cite{brixner2001femtosecond}, as both domains are linked via a Fourier transform. 

Similarly as before, we can extend the initial arbitrary polarization state \eqref{eqA:pol_state} by correlating it to two orthogonal spectral mode functions $F_{n}(\omega)$ such that we obtain a hybrid frequency-polarization state
\begin{eqnarray}
\ket{\Psi}_F=\cos\left(\frac{\theta_F}{2}\right)F_{1}(\omega)\ket{R} + \sin\left(\frac{\theta_F}{2}\right) e^{i\phi_F} F_{2}(\omega) \ket{L},    
\label{eqA:SpeVec_state}
\end{eqnarray}
with $\omega$ being the angular frequency of the light field, $\theta_F\in[0,\pi]$, and $\phi_F\in[0,2\pi]$. Again, a higher-order PS can be constructed, where the poles are given by two orthogonal polarization states, each with a different orthogonal spectral mode. The other states on the surface of the sphere are described by spectral vector fields, where the polarization state varies across the frequency spectrum.

To improve the understanding of such states, it is instructive to discuss them in terms of two specific orthogonal frequency mode functions, such as Hermite Gaussian (HG) spectral modes \cite{brecht2015photon}
\begin{eqnarray}
F_{n}(\omega) = G(\omega) H_n(\omega), 
\end{eqnarray}
where $G(\omega)$ describes a Gaussian profile and $H_n(\omega)$ are the Hermite polynomials. For later convenience, we can introduce the spectrum-scaled frequency, $\Omega=\sqrt{2(\omega_0-\omega)/\sigma}$, where $\omega_0$ is the central frequency of the pulse and $\sigma$ its spectral bandwidth. With $F_1(\omega)=H_0(\omega)=1$, $F_2(\omega)=H_1(\omega)=2\omega$, and the definition of $\Omega$ given above, one can then rewrite Eq.~\eqref{eqA:SpeVec_state} to 
\begin{eqnarray} \label{eqA:SpeVec_state_HG}
\ket{\Psi}_F=G(\Omega) \left[\frac{1}{\Omega}\cos\left(\frac{\theta_F}{2}\right)\ket{R} + \Omega \sin\left(\frac{\theta_F}{2}\right) e^{i\phi_F} \ket{L}\right],   
\end{eqnarray}
where $G(\Omega)=\Omega\,G(\omega)$.
Note again, that for $F_1=F_2$ or monochromatic light fields (plane waves), the higher-order PS reduces to the fundamental one. 

In contrast to spatial vector beams constructed with twisted light fields, these hybrid polarization-frequency states have a different connection to the fundamental PS.
Any point on the higher-order spectral PS contains all polarization states of a meridian (vertical circle) on the fundamental PS, rather than along a parallel line as for their spatial counterpart described in Sec.~\ref{sec:2}.
This feature is demonstrated by Eq.~\eqref{eqA:SpeVec_state_HG} where a point on the higher-order PS is described by one combination of $\theta_F$ and $\phi_F$.
The resulting spectral spectral vector beam contains a polarization state that is changing across the frequency spectrum from $\ket{L}$ at $\omega\rightarrow-\infty$ to $\ket{R}$ at $\omega=\omega_0$, and back again to $\ket{L}$ at $\omega\rightarrow\infty$. 
Hence, the polarization state consists of all points along a meridian on the fundamental PS.

\begin{figure}[ht] 
    \centering
    \includegraphics[width=0.8\linewidth]{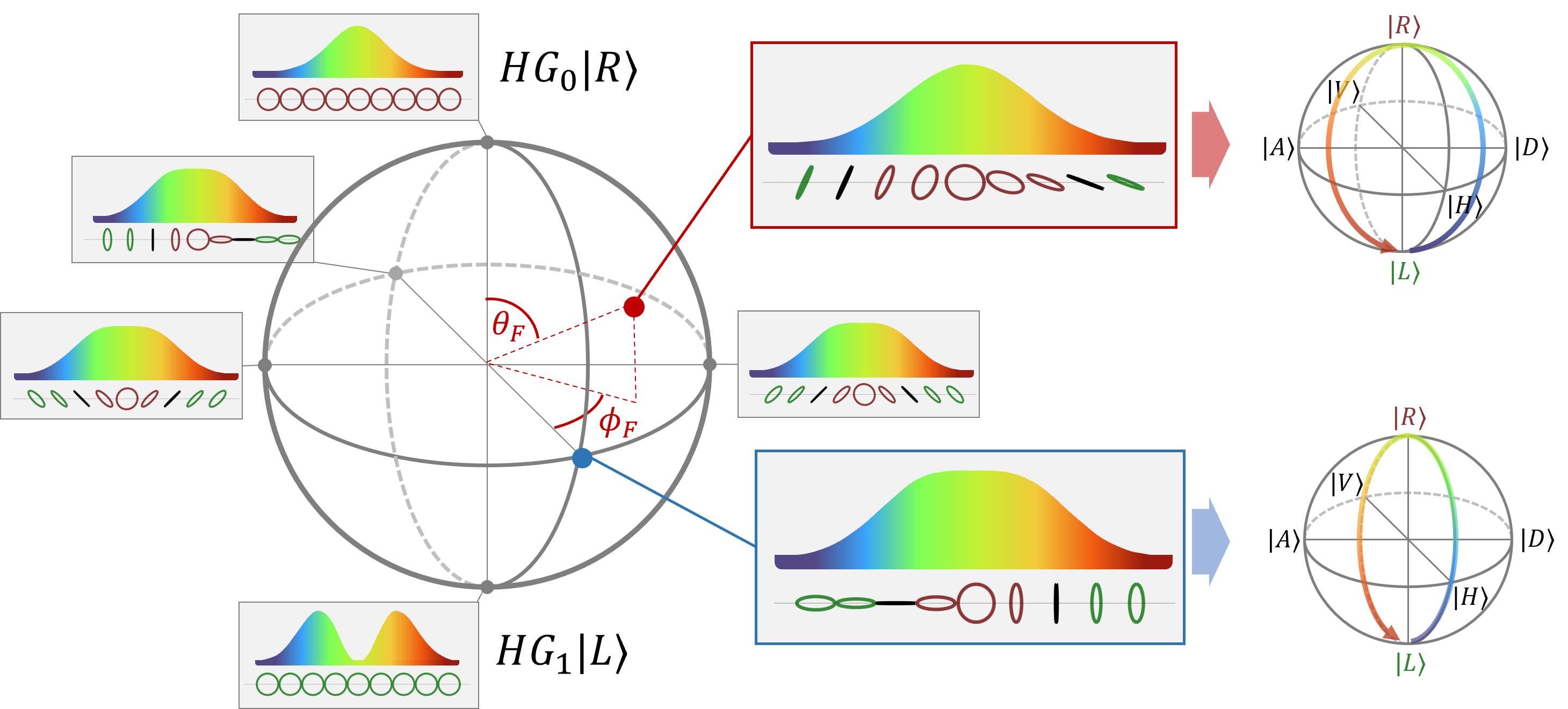}
    \caption{Higher-order Poincaré Sphere (PS) for spectral vector beams. Every point of the higher-order PS corresponds to a meridian on the fundamental PS, as shown here for two examples. The upper example, highlighted in red, is shown for $\theta_F=2\pi/3$ and $\phi_F=\pi/4$, while the lower example (highlighted in blue) shows a state on the equator with $\theta_F=\pi/2$ and $\phi_F=0$.}
    \label{figS:0b}
\end{figure}

The Stokes parameters now have the following form
\begin{subequations}
\begin{align}
S_0 &=|G(\omega)|^2\,\left[\cos^2\left(\frac{\theta_F}{2}\right)+|\Omega|^4\sin^2\left(\frac{\theta_F}{2}\right)\right],\\
S_1/S_0&=4\cos(\phi_F)\,\left[\frac{|\Omega|^2\,\tan\left(\frac{\theta_F}{2}\right)}{1+|\Omega|^4\tan^2\left(\frac{\theta_F}{2}\right)}\right],\\
S_2/S_0&=4\sin(\phi_F)\,\left[\frac{|\Omega|^2\,\tan\left(\frac{\theta_F}{2}\right)}{1+|\Omega|^4\tan^2\left(\frac{\theta_F}{2}\right)}\right],\\
S_3/S_0&=\frac{1-|\Omega|^4\tan^2\left(\frac{\theta_F}{2}\right)}{1+|\Omega|^4\tan^2\left(\frac{\theta_F}{2}\right)}.
\end{align}
\end{subequations} 
If we then define, in analogy to Ref.~\cite{alonsoFullPB}, $\xi=|\Omega|^2\tan\left(\frac{\theta_F}{2}\right)\equiv\tan(\theta/2)$, we can rewrite the Stokes parameters as follows
\begin{subequations}
\begin{align}
S_1/S_0&=\frac{4\xi\cos(\phi_F)}{1+\xi^2}\equiv\cos(\theta)\cos(\varphi),\\
S_2/S_0&=\frac{4\xi\sin(\phi_F)}{1+\xi^2}\equiv-\sin(\theta)\sin(\varphi),\\
S_3/S_0&=\frac{1-\xi^2}{1+\xi^2}\equiv\cos(\theta),
\end{align}
\end{subequations}
where $\{\theta,\varphi=-\phi_F\}$ are the polar and azimuthal angles on the fundamental PS, with the polar angle measured from the equator rather than the North pole. 
Using this convention, we can interpret the triplet $\{S_1/S_0,S_2/S_0,S_3/S_0\}$ as a stereographic projection of a sphere from the South pole. Notice, moreover, that this definition implies that $\theta\equiv\theta(\Omega)=2\arctan\left[|\Omega|^2\,\tan\left(\frac{\theta_F}{2}\right)\right]$ and therefore $\lim_{\Omega\rightarrow\pm\infty}\theta(\Omega)=\pm\pi$. 
This means, that for $\Omega\rightarrow-\infty$, (i.e., at the trailing edge of the pulse) $\theta=-\pi$ and the polarisation of the beam is $\ket{L}$. 
Then, when $\Omega=0$ (meaning $\omega=\omega_0$, i.e., at the carrier frequency) we have $\theta=0$ and the polarisation of the field is $\ket{R}$, as the point representing the pulse on the PS reaches the North pole from the $\theta\in[-\pi,0]$ sector of the sphere. 
Finally, when $\Omega\rightarrow\infty$ (i.e., at the leading edge of the pulse) the polar angle is $\theta=\pi$ and the polarisation of the field returns to $\ket{L}$, i.e., the field reaches the South pole from the $\theta\in[0,\pi]$ sector of the sphere. 
The higher-order PS for spectral modes along with two example points and their relation to the fundamental PS are shown in Fig. \ref{figS:0b}.

\subsection{Higher-Order Poincaré sphere in space and frequency (time)}
\label{sec:4}
\noindent After establishing PSs as a visual representation of correlated polarization and spatial or spectral modes, it is now possible to go beyond this description and look at polarization in connection to both spatial and spectral domains. One can, in fact, simply combine the two descriptions introduced above, such that
\begin{eqnarray}
\ket{\Psi}_{TF}=\cos\left(\frac{\theta_{TF}}{2}\right) T_1(\Vec{r})F_{1}(\omega)\ket{R} + \sin\left(\frac{\theta_{TF}}{2}\right) e^{i\phi_F} T_2(\Vec{r})F_{2}(\omega) \ket{L}.    
\label{eqA:SpaSpeVec_state}
\end{eqnarray}
When using spatial modes carrying $\pm1\hbar$ OAM quanta and the two lowest order HG spectral modes, Eq.~\eqref{eqA:SpaSpeVec_state} can be rewritten as
\begin{eqnarray}  
\ket{\Psi}_{TF}&=&G(r)e^{-i\varphi}G(\omega)\Omega\left[\frac{1}{\Omega}\cos\left(\frac{\theta_{TF}}{2}\right)\ket{R} + \Omega\sin\left(\frac{\theta_{TF}}{2}\right) e^{i(\phi_{TF}+2\varphi)}\ket{L}\right].    
\label{eqA:SpaSpeVec_State_LGHG}
\end{eqnarray}
Again, note that for $T_1=T_2$ and $F_1=F_2$ (monochromatic plane waves), the higher-order PS reduces to the fundamental one. 
If only $T_1=T_2$ ($F_1=F_2$) applies, then the reduction leads to the spectral (spatial) higher order PS. 
The most interesting feature, however, is that light fields described by the state given in Eq.~\eqref{eqA:SpaSpeVec_State_LGHG} include all possible polarization states of the PS.
These states are commonly known as Poincaré beams \cite{alonsoFullPB}.

In contrast to earlier discussions on Poincaré beams, the type of beams described by Eq.~\eqref{eqA:SpaSpeVec_State_LGHG} include all polarization states distributed over the whole spatial-spectral domain, hence the name \emph{Spatio-Spectral Poincaré beams} (SSPB).
Analog to for spatial and spectral vector beams individually, one can see that for SSPBs  the amplitude between left- and right-circular polarization varies across the frequency spectrum, while the relative phase changes across the transverse spatial extend. 

The Stokes parameters in this case are given by
\begin{subequations}
\begin{align}
S_0 &=|G(r)|^2|G(\omega)|^2\,\left[\cos^2\left(\frac{\theta_{TF}}{2}\right)+|\Omega|^4\sin^2\left(\frac{\theta_{TF}}{2}\right)\right],\\
S_1/S_0&=4\cos(\phi_{TF}+2\varphi)\,\left[\frac{|\Omega|^2\,\tan\left(\frac{\theta_{TF}}{2}\right)}{1+|\Omega|^4\tan^2\left(\frac{\theta_{TF}}{2}\right)}\right],\\
S_2/S_0&=4\sin(\phi_{TF}+2\varphi)\,\left[\frac{|\Omega|^2\,\tan\left(\frac{\theta_{TF}}{2}\right)}{1+|\Omega|^4\tan^2\left(\frac{\theta_{TF}}{2}\right)}\right],\\
S_3/S_0&=\frac{1-|\Omega|^4\tan^2\left(\frac{\theta_{TF}}{2}\right)}{1+|\Omega|^4\tan^2\left(\frac{\theta_{TF}}{2}\right)},\\
\end{align}
\end{subequations} 
which we can rewrite using $\xi=|\Omega|^2\tan\left(\frac{\theta_{TF}}{2}\right)\equiv\tan(\theta/2)$ to
\begin{subequations}
\begin{align}
S_1/S_0&=\frac{4\xi\cos(\phi_{TF}+2\varphi)}{1+\xi^2}\equiv2\sin(\theta)\cos(\Phi),\\
S_2/S_0&=\frac{4\xi\sin(\phi_{TF}+2\varphi)}{1+\xi^2}\equiv-2\sin(\theta)\sin(\Phi),\\
S_3/S_0&=\frac{1-\xi^2}{1+\xi^2}\equiv\cos(\theta).
\end{align}
\end{subequations}
This again corresponds to a stereographic projection of a sphere from the North pole, which can be brought back to its original state via the mapping $\Phi=\phi_{TF}+2\varphi$ and $\theta=2\arctan\left[|\Omega|^2\,\tan\left(\frac{\theta_F}{2}\right)\right]$, such that the back-projection of the stereographic map onto the sphere gives the point $P=\{\sin(\theta)\cos(\Phi),\sin(\theta)\sin(\Phi),\cos(\theta)\}$.

If we now take as an example the state on the $S_1/S_0$-axis on the spatio-spectral PS with $\{\theta_{TF}=\pi/2,\phi_{TF}=0\}$, the corresponding point $P$ on the ``normal" PS is given by
\begin{equation}
P_{\pi/2,0}=[\sin(\arctan\Omega)]\{\cos(2\ell\varphi),\sin(2\varphi),0\}.
\end{equation}
For a fixed valued of $\Omega$, the term $\sin(\arctan\Omega)$ is just a numerical prefactor.
$P_{\pi/2,0}$ describes a doubly covered great circle on the sphere at the equator as the spatial coordinate $\varphi$ changes from $0$ to $2\pi$, i.e., the corresponding beam contains all polarisation states on the equator. 
If we instead have a constant value of $\varphi$ and let $\Omega$ vary from $-\infty$ to $+\infty$, i.e., we sweep across the whole spectrum of the beam, $P_{\pi/2,0}$ describes a great circle in the meridian plane, i.e., from the South pole to the North pole. 
Notice, that as $\omega\rightarrow\pm\infty$, $\sin(\arctan\Omega)\rightarrow\pm\pi/2$, which implies that in the second case $\theta\rightarrow\pi/2-\theta$, i.e., the polar angle is measured from the equator instead of from the North pole.
%

\subsection{Experimental realization of a spatio-spectral Poincaré beam}
\label{sec:5}
\begin{figure}[ht] 
    \centering
    \includegraphics[width=0.5\linewidth]{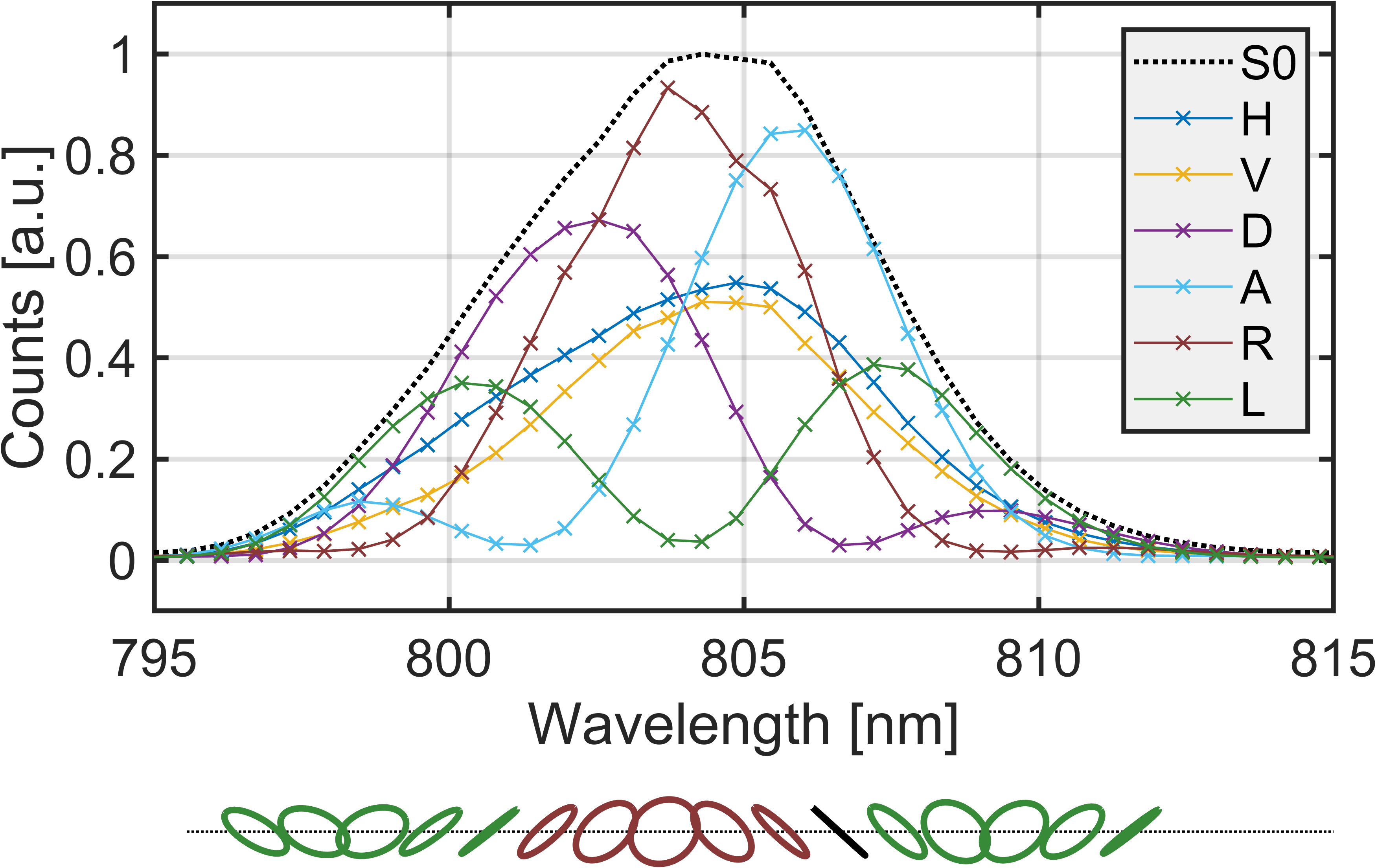}
    \caption{Experimentally measured spectral vector beam. Top: Recorded spectra for different polarizations (data points connected by solid lines) and overall spectral distribution of the whole laser pulse (dotted line).
    \(S_0\) is obtained by averaging over the three possible measurements (see Eq.~\eqref{eq:s0}), each done by adding up the intensities in each basis.
    Bottom: Reconstructed wavelength-dependent polarization pattern.} 
    \label{fig:S_1}
\end{figure}

\noindent As described in the main text, we implement a spatio-spectral Poincaré beam with a simple setup with which we approximate the states described above.
The initial state of our laser is described by a horizontally polarized beam with a Gaussian spectrum $G(\lambda)$ over its wavelength with a bandwidth of approximately 10\,nm centered at 804.5\,nm. 
The spatial profile is approximately described by a Gaussian distribution $G(r)$ of around 2-3\,mm beam waist.
Thus, the initial state can be described by
\begin{eqnarray}
\ket{\Psi}^{exp}&=&G(r)\,G(\lambda)\ket{H}.
\end{eqnarray}
Using a half-wave plate we adjust the polarization such that it has a 45$^\circ$ angle with respect to the optical axis of a subsequent birefringence crystal.
We adjusted the optical axis of the crystal to be aligned with the horizontal polarization, such that with respect to the crystal the state can now be described as 
\begin{eqnarray}
\ket{\Psi}^{exp}&=&\frac{1}{\sqrt{2}}G(r)\,G(\lambda)\left(\ket{H}+\ket{V}\right).
\label{eqA:diagPol}
\end{eqnarray}
Due to the birefringence of the crystal, the vertical components of the pulse experience a time delay $\tau$ with respect to the horizontal ones.
Here, we use a BaB$_2$O$_4$ crystal (BBO) with a cut-angle of 23.4$^\circ$, such that the two parts are temporally split by approximately 220\,fs, which corresponds to the initial pulse length.
This delay in time corresponds to a constant phase gradient over wavelength \cite{kopf2021spectral}, such that we can write 
\begin{eqnarray}
\ket{\Psi}^{exp}_F&=&\frac{1}{\sqrt{2}}G(r)\,G(\lambda)\left(\ket{H}+e^{2i\kappa}\ket{V}\right),
\end{eqnarray}
where $2\kappa= \pi\tau c/\lambda$. The state above can be conveniently rewritten in the circular basis as
\begin{eqnarray}
\ket{\Psi}^{exp}_F&=&G(r)\,G(\lambda)e^{i\kappa}\left[\cos(\kappa)\ket{R}-i\sin(\kappa)\ket{L}\right].
\label{eqA:SpecLab}
\end{eqnarray}
For the wavelength range where $-\pi/2\leq\kappa\leq\pi/2$ applies, we find that the polarization oscillates from left-circular to diagonal, to right-circular, to anti-diagonal, and back to left-circular again.
Hence, we obtain a state that resembles the spectral vector beams described in Eq.~\eqref{eqA:SpeVec_state_HG} and the main text by Eq.~(5) for $\theta_F=\pi/2$ and $\phi_F=\pi/2$.
To verify the introduced spectral vector beam, we perform spectrally-resolved polarization tomography using a set of half- and quarter-wave plates, a polarizing beam splitter, and a spectrometer.
The obtain the spectra for each polarization along with the reconstructed wavelength-dependent polarization pattern as shown in Fig. \ref{fig:S_1}.

\noindent To shape the polarization in space, i.e., generate a spatial vector beam, we use a so-called S-vortex plate, which is a half-wave plate where the optical axis changes along the azimuthal angle $\varphi$.
Thus, upon transmission through the S-vortex plate, circularly polarized light not only flips its polarization but also experiences an azimuthal phase gradient depending on the initial polarization of the light.
\begin{figure}[hb] 
    \centering
    \includegraphics[width=0.95\linewidth]{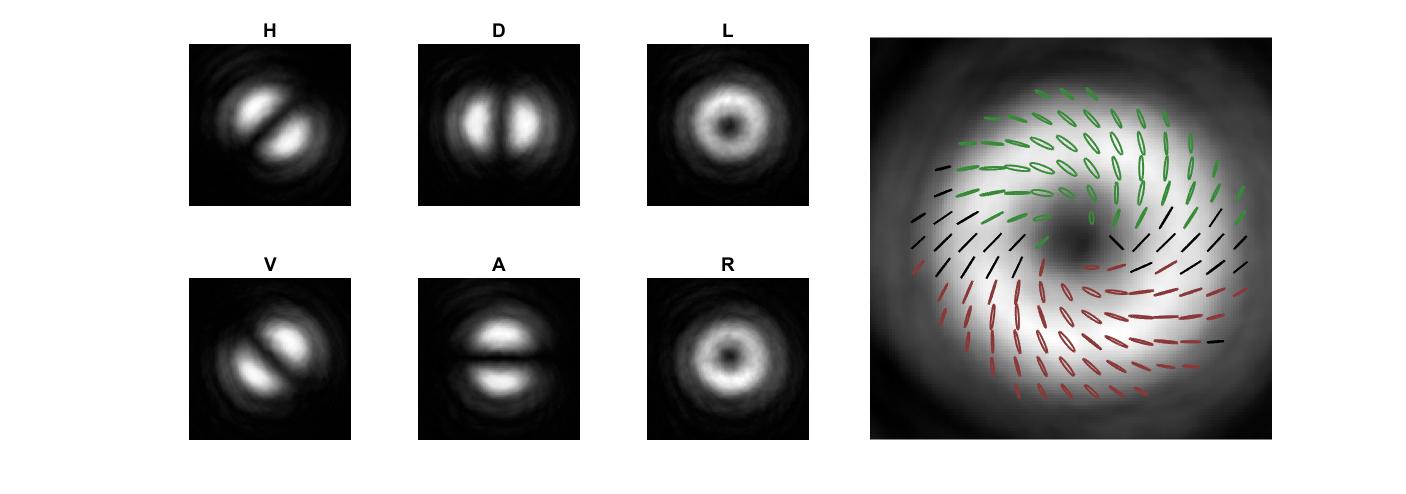}
    \caption{Experimentally measured spatial vector beam. Left: Recorded intensity images for different polarizations. Right:  Reconstructed spatially varying polarization pattern. The intensity in the background shows the averaged $S_0$ intensity, i.e. the average over the sum of the intensity images in each basis.} 
    \label{figS:2}
\end{figure}
Here, the S-vortex plate imprints an azimuthal phase gradient up to 2$\pi$ such that we can replace $\ket{R}$ with $e^{-i\varphi}\ket{L}$ and $\ket{L}$ with $e^{i\varphi}\ket{R}$.
First, we use the S-vortex plate on the laser beam without temporal shaping and reconstruct the resulting polarization pattern in space.
We have a diagonal polarization state as described in Eq.~\eqref{eqA:diagPol}, which can be expressed in terms of circular polarizations as
\begin{eqnarray}
\ket{\Psi}^{exp}&=&\frac{1}{\sqrt{2}}G(r)\,G(\lambda)\left(\ket{R}-i\ket{L}\right).
\end{eqnarray}
After the S-vortex plate, the state is
\begin{eqnarray}
\ket{\Psi}^{exp}_T&=&\frac{1}{\sqrt{2}}G(r)\,G(\lambda)\left(e^{-i\varphi}\ket{L}-ie^{i\varphi}\ket{R}\right),
\end{eqnarray}
which is resembles the state described in Eq.~\eqref{eqA:SpaVec_state_LG} and the main text in Eq.~(3) for $\theta_T=\pi/2$ and $\phi_T=\pi/2$.
The resulting polarization pattern has an azimuthally varying linear polarization with a spiral structure, as shown in the main text in Fig.~1 a), where different spatial vector beams span the higher-order PS.
We reconstruct the polarization pattern with spatially-resolved polarization tomography using a set of half- and quarter-wave plates, a polarizing beam splitter, and a CMOS camera.
All images obtained for different polarizations and the reconstructed polarization patterns are shown in Fig.~\ref{figS:2}.
The patterns nicely match the expected spatial vector beams, with slight imperfections on the lower and upper part of the beam due to experimental imperfections, which is for example visible by comparing the projection on left and right circularly polarized light in Fig.~\ref{figS:2}.

\begin{figure}[ht] 
    \centering
    \includegraphics[width=0.8\linewidth]{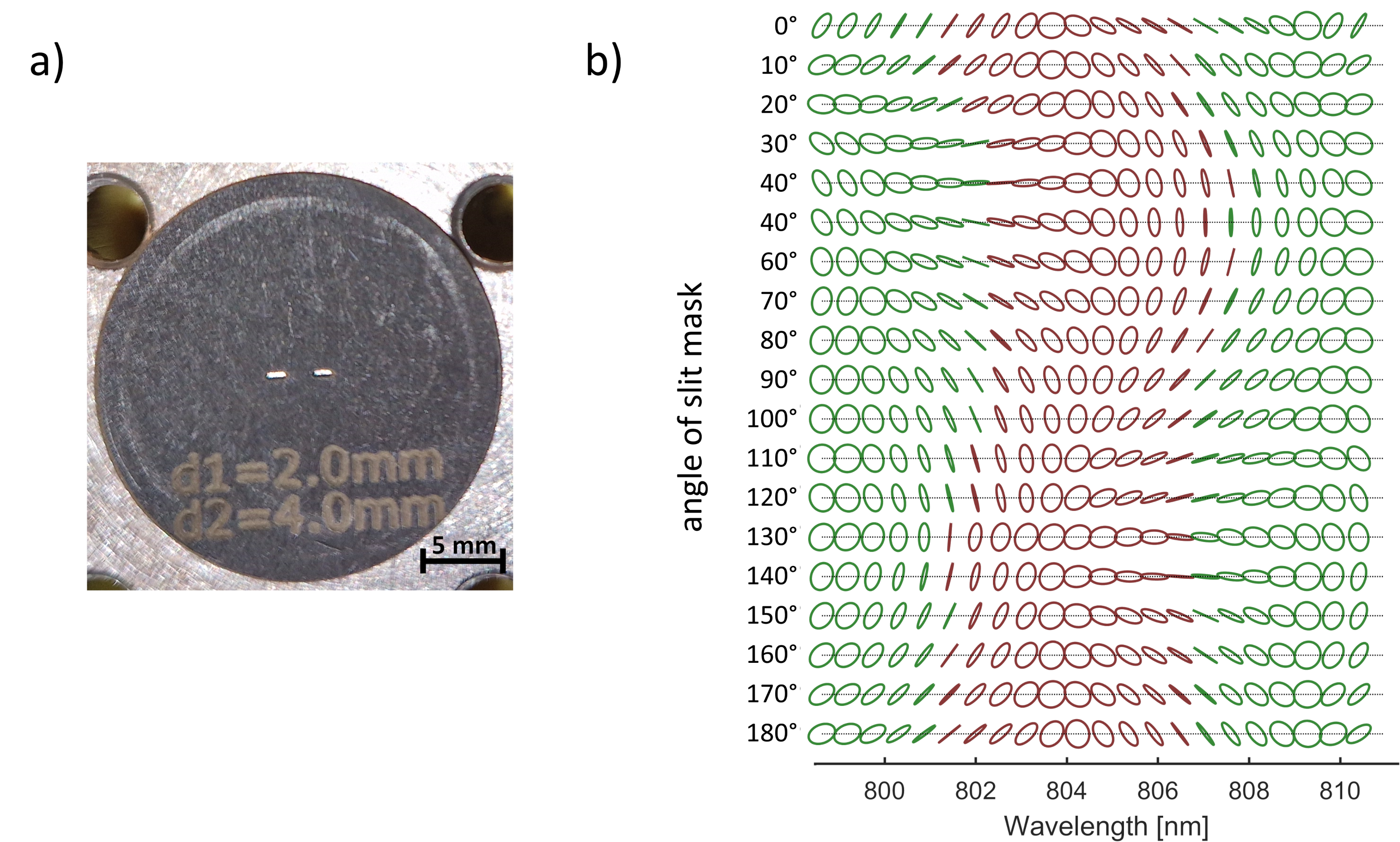}
    \caption{Experimental details and results.
    a) A photograph of the utilized 1-inch slit mask, cut out of black cardboard with a laser cutter. The slits have a measured width of approximately 340\,µm and a length of around 1.10\,mm. The distance between the slits is around 1.65\,mm.
    b) Reconstructed polarization ellipses depending on the angle of the slit mask and wavelength which are measured with the setup described in the text. All reconstructed ellipses are used in Fig.~4 b) of the main text and plotted with their respective positions on the fundamental PS. Right- and left-handed polarizations are depicted by red and green ellipses, respectively. } 
    \label{figS:3}
\end{figure}

After having demonstrated the two techniques to generate spectral and spatial polarization patterns, we combine the two and generate a spatio-spectral vector beam which contains all polarizations, i.e. a spatio-spectral Poincaré beam.
For this, we cascade the birefringent crystal with the S-vortex plate, such that the state of Eq.~\eqref{eqA:SpecLab} becomes
\begin{eqnarray}
\ket{\Psi}^{exp}_{TF}&=&G(r)\,G(\lambda)e^{i\kappa}\left[\cos(\kappa)e^{-i\varphi}\ket{L}-i\sin(\kappa)e^{i\varphi}\ket{R}\right].
\end{eqnarray}
After one reflection of a mirror, which induces a flip of the circular polarizations and a sign-change in the azimuthal phase gradient, the state can be written as
\begin{eqnarray}
\ket{\Psi}^{exp}_{TF}&=&G(r)\,G(\lambda)e^{i\kappa}\left[\cos(\kappa)e^{i\varphi}\ket{R}-i\sin(\kappa)e^{-i\varphi}\ket{L}\right].
\label{eqa:SpaSpecLab}
\end{eqnarray}
As described in the main text, this state describes the spatio-spectral Poincaré beam of Eq.~\eqref{eqA:SpaSpeVec_State_LGHG} or Eq.~(5) in the main text for $\theta_{TF}=\pi/2$ and $\phi_{TF}=\pi/2$.
To show the resulting polarization pattern, we first filter the beam's polarization using a set of half- and quarter-wave plates and a polarizing beam splitter. 
Subsequently, we filter for a specific angle using the angular slit mask shown in Fig.~\ref{figS:3} a).
The mask is fabricated by laser-cutting two slits into a black card board.
Finally, we use a spectrometer to measure the spectral intensity for different polarization and mask setting, such that we can tomographically reconstruct the spatio-spectral polarization pattern.

To show the resulting polarization pattern, we filter the beam first for a specific angle using the angular slit mask shown in Fig.~\ref{figS:3} a).
The mask is fabricated by laser-cutting two slits into a black card board.
Subsequently, we perform a spectrally-resolved polarization tomography using a set of half- and quarter-wave plates, a polarizing beam splitter, and a spectrometer for different spatial mask rotation angles.
The results are given in Fig.~\ref{figS:3} b) and in the main text in Fig.~4.
The results nicely demonstrate that it is possible to generate a spatio-spectral Poincaré beam using minimally with only two optical elements, a birefringent material and an S-vortex plate (given that the input beam is linearly polarized in respect to the birefringent crystal axis).

\end{document}